\documentclass[twocolumn]{aastex701}

\usepackage{graphicx}
\usepackage{txfonts}
\usepackage{subcaption}
\usepackage{multirow}
\usepackage{caption}
\usepackage{xspace}

\newcommand{\halpha}{H$\alpha$\xspace}
\newcommand{\hbeta}{H$\beta$\xspace}
\newcommand{\oiii}{[\ion{O}{3}]\xspace}
\newcommand{\nii}{[\ion{N}{2}]\xspace}
\newcommand{\sii}{[\ion{S}{2}]\xspace}

\usepackage[dvipsnames]{xcolor}
\definecolor{cfsun}{HTML}{be002f}

\definecolor{cxwang}{HTML}{007839}

\begin{document}

\title{A massive barred spiral galaxy at $z = 5.102$ discovered by JWST}

\author[orcid=0009-0002-8312-1002]{Xiaohan Wang}
% \altaffiliation{Tsinghua University}
\affiliation{Department of Astronomy, Tsinghua University, Beijing 100084, People’s Republic of China}
\affiliation{Department of Astronomy, Westlake University, Hangzhou 310030, Zhejiang Province, People’s Republic of China}
\email[show]{xiaohanw78@gmail.com}  

\author[0000-0002-4622-6617]{Fengwu Sun}
\affiliation{Center for Astrophysics $|$ Harvard \& Smithsonian, 60 Garden St., Cambridge, MA 02138, USA}
\email[show]{fengwu.sun@cfa.harvard.edu}  

\author[orcid=0000-0003-3983-5438]{Yoshihisa Asada}
\affiliation{Dunlap Institute for Astronomy and Astrophysics, 50 St. George Street, Toronto, ON M5S 3H4, Canada}
\email{yoshi.asada@utoronto.ca}

\author[0000-0003-0111-8249]{Yunjing Wu}
\affiliation{Kavli Institute for the Physics and Mathematics of the Universe (WPI), The University of Tokyo Institutes for Advanced Study, The University of Tokyo, Kashiwa, Chiba 277-8583, Japan}
\affiliation{Center for Data-Driven Discovery, Kavli IPMU (WPI), UTIAS, The University of Tokyo, Kashiwa, Chiba 277-8583, Japan}
\email{yunjing.wu@ipmu.jp}

\author[0000-0001-6052-4234]{Xiaojing Lin}
\affiliation{Department of Astronomy, Tsinghua University, Beijing 100084, People’s Republic of China}
\email{xiaojinglin.astro@gmail.com}

\author[orcid=0000-0001-8317-2788]{Shude Mao}
\affiliation{Department of Astronomy, Westlake University, Hangzhou 310030, Zhejiang Province, People’s Republic of China}
\email{shude.mao@westlake.edu.cn}

%%%%%%%%%%%%%%%%%%%%%%%%%%%%%%%%%%%%
%%% Addd your author info below: %%%
%%%%%%%%%%%%%%%%%%%%%%%%%%%%%%%%%%%%

\author[0009-0003-4133-0292]{Sijia Cai}
\affiliation{Department of Astronomy, Tsinghua University, Beijing 100084, People’s Republic of China}
\email{caisj23@mails.tsinghua.edu.cn}

\author[0000-0001-8467-6478]{Zheng Cai}
\affiliation{Department of Astronomy, Tsinghua University, Beijing 100084, People’s Republic of China}
\email{zcai@mail.tsinghua.edu.cn}

\author[0000-0002-3805-0789]{Chian-Chou Chen}
\affiliation{Academia Sinica Institute of Astronomy and Astrophysics (ASIAA), No. 1, Sec. 4, Roosevelt Road, Taipei 106319, Taiwan}
\affiliation{East Asian Observatory, 660 N. A'ohoku Pl., Hilo, HI 96720, USA}
\email{ccchen@asiaa.sinica.edu.tw}

\author[0000-0003-1060-0723]{Wenlei Chen}
\affiliation{Department of Physics, Oklahoma State University, 145 Physical Sciences Bldg, Stillwater, OK 74078, USA}
\email{wenlei.chen@okstate.edu}

\author[0000-0003-0202-0534]{Cheng Cheng}
\affiliation{ Chinese Academy of Sciences South America Center for Astronomy, National Astronomical Observatories, Chinese Academy of Sciences, Beijing 100101, People’s Republic of China}
\affiliation{Key Laboratory of Optical Astronomy, NAOC, 20A Datun Road, Chaoyang District, Beijing 100101, People’s Republic of China}
\email{chengcheng@nao.cas.cn}

\author[0000-0003-1949-7638]{Christopher J. Conselice}
\affiliation{Jodrell Bank Centre for Astrophysics, Alan Turing Building, University of Manchester, Oxford Road, Manchester M13 9PL, UK}
\email{conselice@gmail.com}

\author[0000-0003-0348-2917]{Miroslava Dessauges-Zavadsky}
\affiliation{Department of Astronomy, University of Geneva, Chemin Pegasi 51, 1290 Versoix, Switzerland}
\email{Miroslava.Dessauges@unige.ch}

\author[0000-0002-8726-7685]{Daneil Espada}
\affiliation{Departamento de Física Teórica y del Cosmos, Campus de Fuentenueva, Edificio Mecenas, Universidad de Granada, E-18071, Granada, Spain}
\affiliation{Instituto Carlos I de Física Teórica y Computacional, Facultad de Ciencias, E-18071, Granada, Spain}
\email{despada@ugr.es}

\author[0000-0003-1625-8009]{Brenda L. Frye}
\affiliation{Department of Astronomy/Steward Observatory, 933 N. Cherry Ave, Tucson, AZ 85721, USA}
\email{bfrye@arizona.edu}

\author[0000-0002-3405-5646]{Jean-Baptiste Jolly}
\affiliation{Max-Planck-Institut für extraterrestrische Physik, 85748 Garching, Germany}
\email{jbjolly@mpe.mpg.de}

\author[0000-0002-6610-2048]{Anton M. Koekemoer}
\affiliation{Space Telescope Science Institute, 3700 San Martin Drive, Baltimore, MD 21218, USA}
\email{koekemoer@stsci.edu}

\author[0000-0002-4052-2394]{Kotaro Kohno}
\affiliation{Institute of Astronomy, Graduate School of Science, The University of Tokyo, 2-21-1 Osawa, Mitaka, Tokyo 181-0015, Japan}
\affiliation{Research Center for the Early Universe, Graduate School of Science, The University of Tokyo, 7-3-1 Hongo, Bunkyo-ku, Tokyo 113-0033, Japan}
\email{kkohno@ioa.s.u-tokyo.ac.jp}

\author[0000-0001-6251-649X]{Mingyu Li}
\affiliation{Department of Astronomy, Tsinghua University, Beijing 100084, People’s Republic of China}
\affiliation{Kavli Institute for Cosmology, University of Cambridge, Madingley Road, Cambridge CB3 0HA, UK}
\affiliation{Cavendish Laboratory, University of Cambridge, 19 JJ Thomson Avenue, Cambridge CB3 0HE, UK}
\email{lmytime@hotmail.com}

\author[0000-0003-3243-9969]{Nicholas Martis}
\affiliation{Faculty of Mathematics and Physics, Jadranska ulica 19, SI-1000 Ljubljana, Slovenia}
\email{nicholas.martis@fmf.uni-lj.si}

\author[orcid=0000-0003-1937-0573,gname=Hideki,sname=Umehata]{Hideki Umehata}
\affiliation{Institute for Advanced Research, Nagoya University, Furocho, Chikusa, Nagoya 464-8602, Japan}
\affiliation{Department of Physics, Graduate School of Science, Nagoya University, Furocho, Chikusa, Nagoya 464-8602, Japan}
\email{hideki.umehata@gmail.com}

\author[0000-0002-9593-8274]{Weichen Wang}
\affiliation{Dipartimento di Fisica, Università degli Studi di Milano-Bicocca, Piazza della Scienza 3, 20126 Milano, Italy}
\email{weichen.wang@unimib.it}

\author[0000-0001-8156-6281]{Rogier A. Windhorst}
\affiliation{School of Earth and Space Exploration, Arizona State University, Tempe, AZ 85287-6004, USA}
\email{Rogier.Windhorst@asu.edu}

\author[0000-0002-0350-4488]{Adi Zitrin}
\affiliation{Department of Physics, Ben-Gurion University of the Negev, P.O. Box 653, Be’er-Sheva 84105, Israel}
\email{adizitrin@gmail.com}

\begin{abstract}
We report M1149-BSG-z5, a
% candidate for the earliest
barred spiral galaxy at
% a redshift of 
$z = 5.102$, identified in the parallel field of MACS J1149+2223 with JWST and HST.
M1149-BSG-z5 is the highest redshift barred galaxy candidate to date.
Both isophote ellipse fitting and structural modeling support a stellar bar of length $a_\mathrm{bar} \approx 4.5$\,kpc, and
extended spiral arms peaking at $r \approx 5.5$\,kpc.
M1149-BSG-z5 is a massive main sequence star-forming galaxy, % on the main sequence, 
with a stellar mass of $ 10^{10.45}\rm M_\odot$ and a star-formation rate of $144\,\rm M_\odot/yr$.
%, residing on the star formation main sequence.
A concentrated bulge is embedded in an extended disk with a global Sérsic index $n = 2.37$.
With an effective radius of $R_{e} = 2.61\rm \ kpc$, M1149-BSG-z5 is larger than typical galaxies at $z \sim 5$ and comparable to barred galaxies at $2 < z < 4$.
% M1149-BSG-z5 also hosts a broad-line AGN; its low black-hole-to-stellar mass ratio, $\rm M_{\rm BH}/M_\ast\sim10^{-3}$, and metal-enriched emission-line properties further point to a chemically evolved host system. 
M1149-BSG-z5 also hosts a broad-line AGN, with a relatively low black-hole-to-stellar mass ratio of $\rm M_{\rm BH}/M_\ast\sim10^{-3}$. Its metal-enriched emission-line properties
% further 
indicate that it
% M1149-BSG-z5 
is already chemically evolved.
These properties imply M1149-BSG-z5 as an early-assembled and structurally evolved galaxy.
% Spatially resolved SED fitting displays asymmetries in star formation and dust attenuation across the bar region, indicating potential bar-driven gas inflows and triggered star formation. The derived spatially resolved properties also indicate that the bar component may have assembled as early as $z \sim 6$.
We also find that M1149-BSG-z5 resides in an overdense region with a nearby companion galaxy, suggesting an interaction-driven bar formation mechanism.
Its concentrated light, early assembly and main-sequence star formation also suggest baryon-dominated, gas-rich conditions, where gravitational instability can further accelerate the bar formation.
% Future spectroscopic confirmation and dynamical constraints will be essential to help understand the mechanisms of early bar formation.

\end{abstract}


\section{Introduction}
\label{Intro}
Bars are important structures in galaxies highly related to galaxy secular evolution.
Bars are common in nearby disks; about 60\% of disk galaxies in the local universe host bar structures in near-infrared observations \citep{Eskridge2000, Knapen2000, Laurikainen2004, Marinova2007, M2007, Sheth2008, Buta2015, Erwin2018}.
Supported by stars on elongated orbits, bars feature a non-axisymmetric potential that drives angular momentum redistribution and radial gas inflows and outflows, related to central structure formation and quenching \citep{Cheung2013, Sheth2005, Fanali2015, Lin2017, Spinoso2017, Fraser-McKelvie2020}.
% Bar structures, supported bar stars in elongated orbits, are highly related to gas inflow and outflows, angular momentum redistribution, central structure formation and quenching \citep{Cheung2013, Sheth2005, Fanali2015, Lin2017, Spinoso2017, Fraser-McKelvie2020}.

Bar structures can arise spontaneously from non-axisymmetric perturbation modes amplified via gravitational instabilities in dynamically cold disks \citep[e.g.][]{Hohl1971, Sellwood1981, Toomre1981, AS1986}; dynamically hot structures can inhibit instabilities and suppress bar formation \citep[e.g.][]{AS1986, Sheth2012}.
Beyond internal secular evolution, bar structures may also be induced by tidal interactions \citep[e.g.][]{Gerin1990, Noguchi1996, Lokas2016, Peschken2019} and wet mergers \citep[e.g.][]{Athanassoula2016}. The external effect on bar formation, however, is highly uncertain and dependent on galaxy mass, orbital parameters, and gas fraction \citep[e.g.][]{MA2012, Casteels2013}.
% The relative phase angle of the companion at pericenter can exert negative torques that weaken an existing bar's strength \citep[e.g.][]{Gerin1990}.
% In gas-rich systems, rapid gas infall or interactions can heat the stellar component and therefore prevent bar formation or weaken bars \citep[e.g.][]{Noguchi1996, Berentzen1998, Berentzen2004}.

At high redshifts, the cosmic environment is drastically different from the local Universe, characterized by high molecular gas fraction \citep{Tacconi2010, Carilli2013}, higher merger and interaction rate \citep{Fakhouri2010, Duan2025, Puskas2025}, more intense star-forming activity \citep{Madau2014, Speagle2014, Looser2025, Rinaldi2025} and stronger feedback \citep{Hopkins2014, Carniani2024}.
High-redshift disk galaxies experience strong perturbations and tend to be dynamically hot \citep{Bournaud2007, Dekel2009, Foster2009, Ceverino2010, Kassin2012, Wisnioski2015}, in which bar structure formation is expected to be suppressed.
Observations from the Hubble Space Telescope (HST) have found consistent results: the bar fractions among disk galaxies drop to $\sim 20\%$ at $z \sim 0.5$ and $\sim 10\%$ at $z \sim 1$ \citep{Sheth2008, Melvin2014, Simmons2014}.
Certain zoom-in cosmological simulations have shown a quantitatively consistent trend with bar fraction approaching almost zero at $z > 2$ \citep[e.g.,][]{Kraljic2012}.

However, with the advent of the James Webb Space Telescope (JWST), recent observations at better sensitivity and finer angular resolution have revealed more bars than previously expected.
The bar fraction derived by JWST observations is about twice that previously measured by HST at $0.5 < z < 1.0$ \citep{LeConte2024, Geron2025, Guo2025, Huertas-Company2025, Espejo_Salcedo2025, LeConte2026}.
Extending constraints into higher redshifts, JWST observations have found that barred galaxies emerge as early as $z \sim 4$, with observed fractions of $3-7\%$ at $z \sim 3.5$ \citep{Geron2025, Guo2025, Huertas-Company2025, LeConte2026}.
Simulation of observations even suggests that these fractions may be underestimated \citep{Liang2024}.
In addition to statistical analysis, well-developed bars have been detected at $1.5 < z < 3.5$ \citep{Costantin2023_bar, Guo2023, Umehata2025, Huang2025, Amvrosiadis2025, Kalita2026}, and galaxies with bar-like structures are reported at $z=4.055$ \citep{Boogaard2026}, $z = 4.260$ \citep{Smail2023} and $z = 4.407$ \citep[][with ALMA]{Tsukui2024}.

This early emergence and abundance of barred galaxies, coupled with the detection of dynamically cold disks at high redshifts \citep[e.g.][]{Rizzo2020, Lelli2021, Robertson2023, Wu2023, Ferreira2023, Wang2025, Jain2025, allingham2026}, challenges the scenario of bar formation in cold disks and the expected epoch of disk settling.
Some simulations have shown consistent results of bar emergence as early as $z \sim 4$ and high abundance of bars at $z > 3$ \citep{Rosas-Guevara2022, Fragkoudi2025}, while some even suggest bar formation as early as $z > 5$ \citep{Lokas2025}.
% Further identification and observational constraints of early bars, as well as pushing the search to higher redshifts for the earliest stellar bars, can help to understand the bar formation mechanisms in the early Universe.
Identifying more early bars and pushing the search to higher redshifts will help constrain the bar-formation mechanisms in the early Universe.

% Auriga, z ~ 3, 20\%, merger-induced
% Bi2022, more gas rich, interaction, 
% TNG50, > 40\% at z > 3, less gas rich, baryon domination, first bar as early as z ~ 4, more baryon dominated, early assembly

% 1. Guo2024, M > 10, z ~ 4, 2-4 < 10\%, emerging at z ~ 4; high frequency of companion galaxies at z >= 1.5, "nearby neighbors or/and signs of tidal interactions", 6.4 ± 2.4 per cent at z ∼ 3.5.
% 2. Le Conte24, 17.8+5.1−4.8 per cent at 1 ≤ z < 2, 13.8+6.5−5.8 at 2 ≤ z ≤ 3
% 0.16+0.03−0.03 at 1 ≤ z < 2; 0.08+0.02−0.01 at 2 ≤ z < 3; 0.07+0.03−0.01 at 3 ≤ z ≤ 4. 
% 3. Geron25 3\% at z 3-4
% 4. MH25, consistent, and even at z 4-5, but nobars at z > 5? no more data.. 
% "Notably, the larger COSMOS-Web survey area allows us to probe higher redshifts than Guo et al. (2025), revealing that the abundance of stellar bars is consistent with zero at z ∼ 4 − 5."

% When do the first bar galaxies emerge? what would happen for these galaxies?

In this work, we report a candidate for one of the earliest known barred spiral galaxies at $z = 5.102$.
This galaxy is observed in the parallel field of MACS\,J1149+2223, hereafter M1149.
We refer to this galaxy as M1149-BSG-z5 (where BSG stands for barred spiral galaxy) in the rest of the paper. 
M1149-BSG-z5 is identified as hosting a bar with both visual classification and ellipse fitting.
We summarize the physical and structural properties of M1149-BSG-z5 in Table~\ref{props}.
We describe the datasets and measurements in Sec.~\ref{Obs}. 
In Sec.~\ref{Results} we present the structural, stellar population and emission-line properties of M1149-BSG-z5.
In Sec.~\ref{Discussion} we discuss potential mechanisms of bar formation at redshift of 5, including interactions and gravitational instability.
We summarize in Sec.~\ref{Conclusions}.
We adopt the AB magnitude system \citep{Oke1983}. We also adopt a flat $\Lambda$CDM cosmology with $H_0 = 67.66\rm \ km\ s^{-1}\ Mpc^{-1}$, $\Omega_m = 0.31$ and $\Omega_{\Lambda} = 0.69$ \citep[][]{Planck2020}.

\section{Observations and Measurements}
\label{Obs}

\subsection{JWST and HST imaging observations}

\begin{figure*}
    \centering
    \begin{subfigure}{0.42\textwidth}
        \centering
        \hspace{1em}
        \includegraphics[width = \columnwidth]{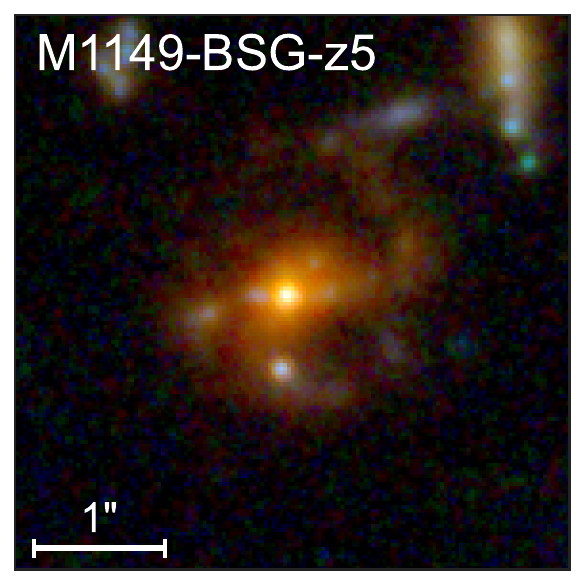}
    \end{subfigure}
    \begin{subfigure}{0.56\textwidth}
        \includegraphics[width = \columnwidth]{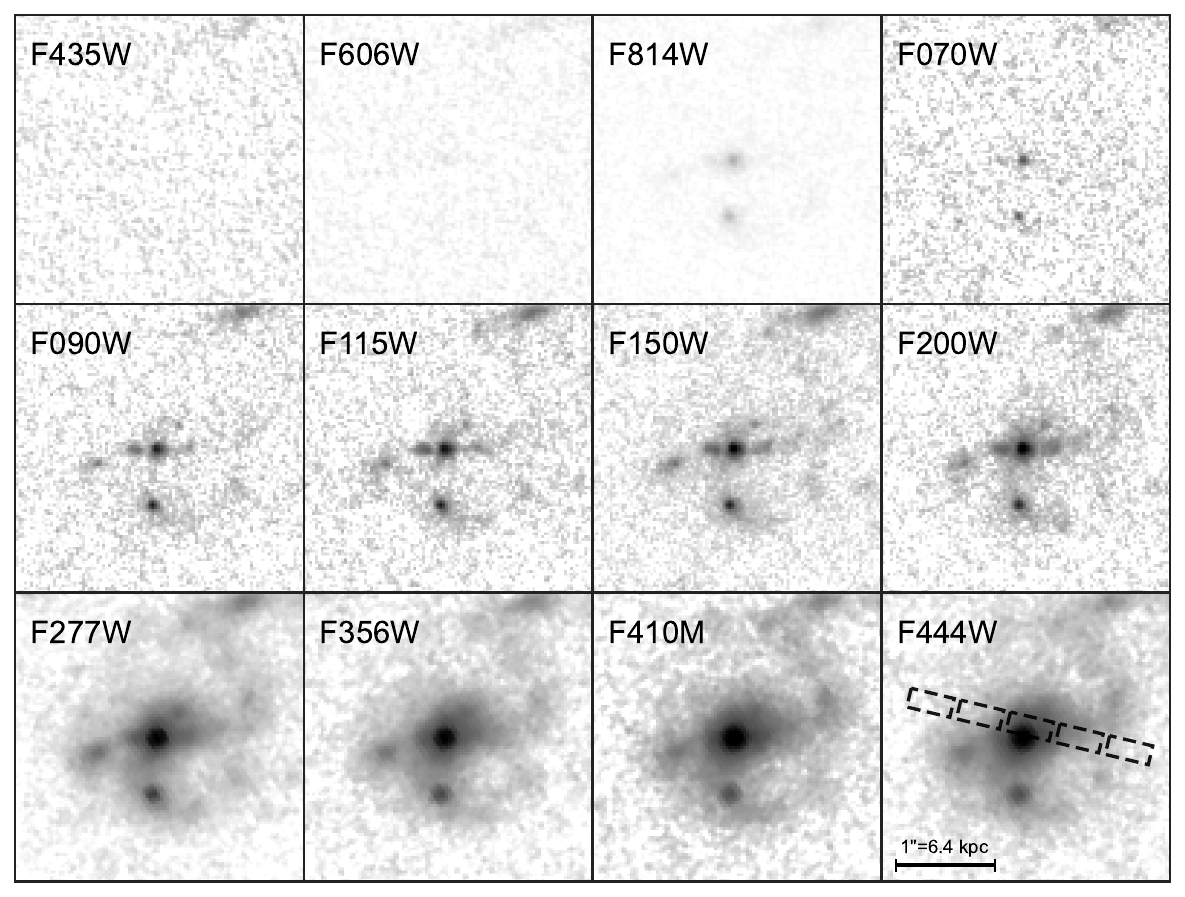}
    \end{subfigure}
    \begin{subfigure}{0.98\textwidth}
        \centering
        \includegraphics[width = \columnwidth]{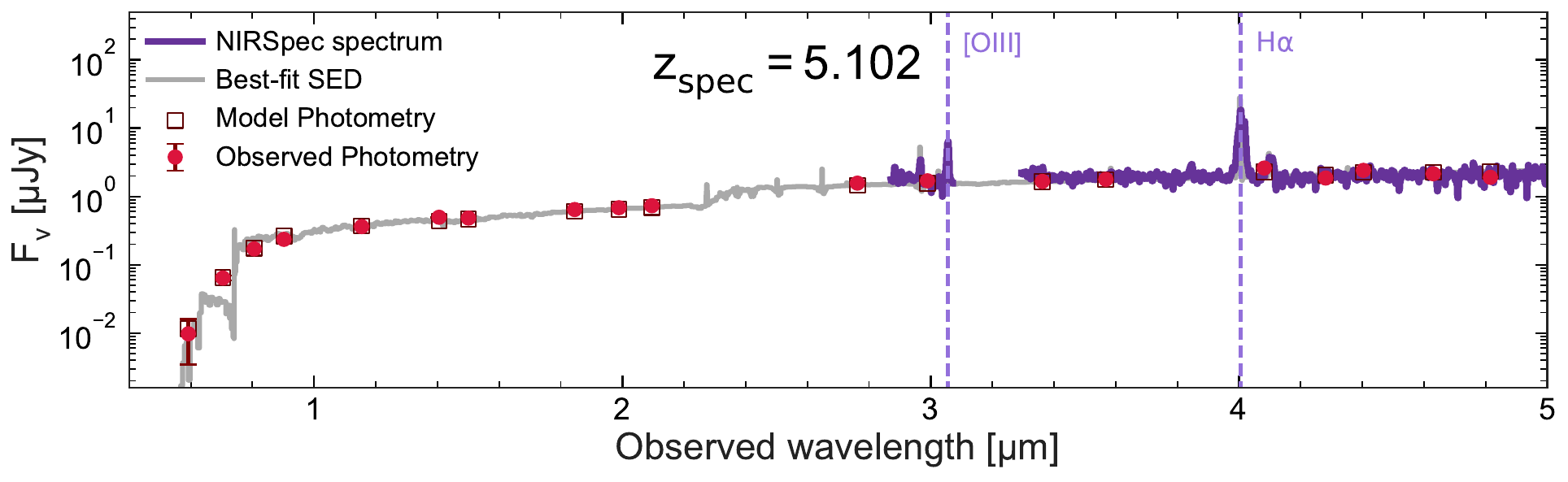}
    \end{subfigure} 
    \begin{subfigure}{0.95\textwidth}
        \centering
        \hspace{5em}
        \includegraphics[width = \columnwidth]{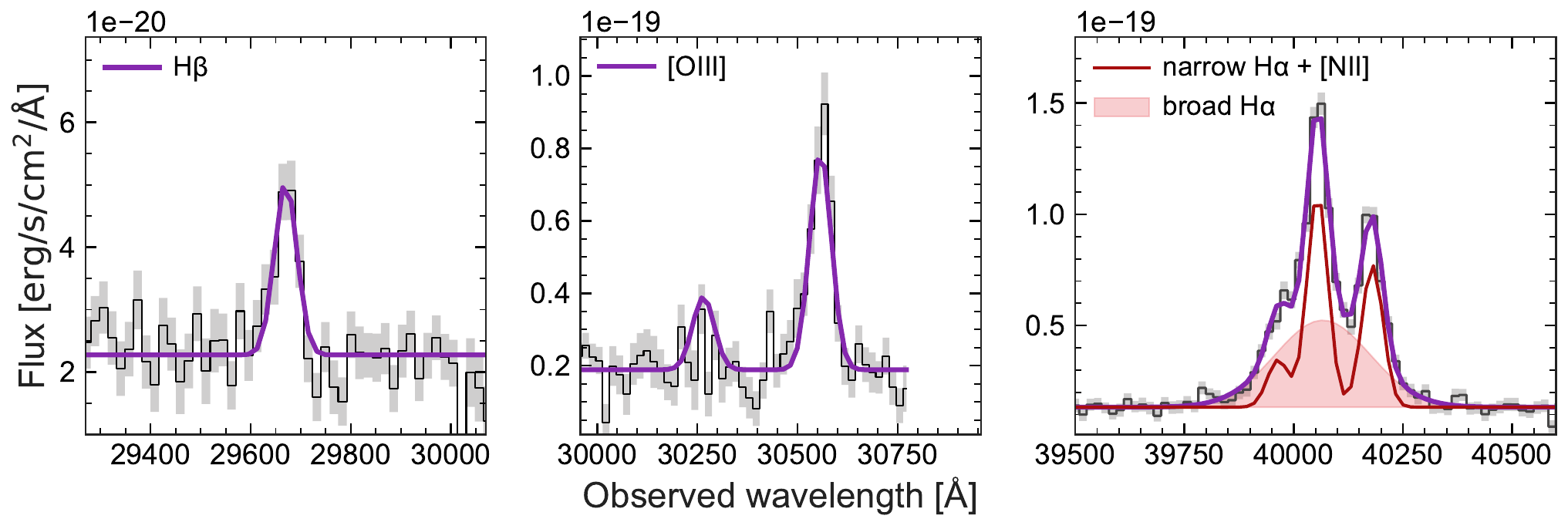}
    \end{subfigure} 
    \caption{Top left: composite false-color image of M1149-BSG-z5, created with JWST imaging data taken in F090W+F115W+F150W (blue), F200W+F277W+F300M (green), and F356W+F410M+F444W (red). Image size is $4\farcs35 \times 4\farcs35$ (north up, east left).
    Top right: HST and JWST images of M1149-BSG-z5 across 12 bands at 0.4--5.0\,\micron. The NIRSpec MSA coverage is shown with dashed squares in the image of F444W.
    Middle panel: best-fit SED model from \texttt{CIGALE}, accompanied with HST and JWST Kron photometry and JWST NIRSpec G395M/F290LP spectrum. The spectrum is scaled to the photometry observations.
    Botton panel, from left to right: spectral regions of main emission lines, H$\beta$, [O\,{\sc iii}] $\lambda\lambda4959,5007$, and H$\alpha$+[N\,{\sc ii}] $\lambda\lambda6548,6583$. Note that the zoom-in spectra are not rescaled.
    Purple curves show the best-fit line models. For H$\alpha$+[N\,{\sc ii}], the narrow H$\alpha$+[N\,{\sc ii}] component is shown in red, with an extra broad H$\alpha$ component shown by the shaded red profile.}
    \label{img}
\end{figure*}

M1149-BSG-z5 was identified in the NIRISS \citep{Doyon23} imaging parallel field of JWST Cycle-2 program \textsl{``Medium-band Astrophysics with the Grism of NIRCam in Frontier Fields''} (MAGNIF, JWST-GO-2883; PI: F.\ Sun).
Four Frontier-Field \citep{Lotz2017} galaxy clusters, namely MACS\,J0416, MACS\,J1149, Abell 370 and Abell 2744, were observed through NIRCam (imaging and grism spectroscopy) as the primary instrument and NIRISS (imaging in the F430M and F480M band) in parallel.
In the M1149 cluster field, archival JWST imaging observations in Cycle 1 and 2 from GTO-1199 (PI: Stiavelli; \citealt{Stiavelli2023}), GTO-1208 (CANUCS; PI: Willott; \citealt{Willott22}), and GO-3362 (Technicolor; PI: Muzzin; \citealt{Sarrouh2026}) were also included for our data analyses. 
The JWST imaging mosaics that we produced include 16 bands (F070W, F090W, F115W, F140M, F182M, F200W, F210M, F277W, F300M, F335M, F356W, F410M, F430M, F444W, F460M and F480M) at 0.6--5.0\,\micron.
The JWST images were reduced and mosaicked to a homogeneous pixel scale of $0\farcs03$ using a customized stage-1/2/3 JWST pipeline  \verb|v1.11.2|  \citep{bushouse23} and CRDS calibration reference file context \verb|jwst_1188.pmap|, detailed in \citet{FuS25}.

The M1149 field benefits from deep HST/ACS imaging observations from the Hubble Frontier Field program \citep{Lotz2017} and the Beyond Ultradeep Frontier Fields And Legacy Observations (BUFFALO) program \citep{Steinhardt2020}, including F435W, F606W and F814W filters.
We directly made use of the HST/ACS imaging data products from the BUFFALO data release.

\begin{table}[]
    \centering
    \caption{Physical Properties of M1149-BSG-z5}
    \begin{tabular}{cc}
        \hline
        Galaxy & M1149-BSG-z5 \\
        \hline
        RA (deg) & 177.39742\\
        DEC (deg) & 22.28442\\
        $z_{\rm spec}$ & $5.1015\pm0.0002$ \\\hline
        $R_e (\rm kpc) $ & $2.61\pm 0.02$ \\
        Sérsic $n$ & $2.37\pm0.03$ \\
        B/T &  $0.42 \pm 0.01$\\
        $R_\mathrm{bar}$ (kpc) & 4.50 \\
        $R_\mathrm{spiral}$ (kpc) & 5.52 \\
        \hline
        $\log \rm [SFR / (M_\odot yr^{-1})], SED$ & $2.16^{+0.07}_{-0.08}$ \\
        $\log (M_*/M_\odot)$ & $10.45\pm{0.05}$ \\
        $A_v$ (mag) & $1.06\pm 0.13$\\
        Mass-weighted Age (Myr) & $148\pm36$\\
        % Bar Age (Myr) &  \fsun{TBD} \\
        \hline
    \end{tabular}
    \label{props}
    \tablecomments{
    % M1149-BSG-z5 is identified in the parallel field of galaxy cluster MACS\, J1149 (6\farcm8 from the brightest cluster galaxy) and therefore the lensing magnification is negligible.
    The structural properties of M1149-BSG-z5 are measured from the stacked F277W, F356W, and F444W image, which corresponds to rest-frame optical wavelengths.}
\end{table} 
% Add notes here

% Customized steps include but not limit to 1/f, wisp and modeled background subtraction.
% 1208, 3362, 2883

\subsection{Photometric catalog and photometric redshifts}
\label{sec_02b:phot}
% \begin{figure}
%     \centering
%     \includegraphics[width = \columnwidth]{figures/sed_pz.pdf}
%     \caption{}
% \end{figure}

The construction of the photometric catalog and inference of photometric redshifts have been detailed by \citet{FuS25}.
First, we stacked the JWST images taken at 1.8--5.0\,\micron\ to produce a detection image. 
We then iteratively median-filtered (sharpened) the detection image to suppress the diffuse light from intracluster light, and to enhance the detection of faint blended sources.
We then extracted and deblended sources using a \texttt{photutils} pipeline \citep{Bradley2022}, and conducted photometry in all bands with various circular and Kron apertures and local background subtraction. 
All photometry was aperture-corrected using a point-source aperture correction factor based on \texttt{STPSF} models \citep{Perrin14}.
% Photometric redshifts adopted in this work were measured using an \texttt{EA}

Photometric redshifts were measured through an \texttt{EAZY} \citep{Brammer2008} pipeline with $r = 0 \farcs 1$ aperture photometry and templates from \citet{Hainline2024} optimized for compact, high-redshift star-forming sources.
We set a redshift range of (0.01, 30) with a step $\Delta z = 0.01 (1 + z)$ and a photometric error floor of 5\%, and adopt the EAZY template error file \verb|template error.v2.0.zfourge|.
No apparent magnitude prior was applied. Throughout this work, we adopt the EAZY maximum-a-posteriori redshift, $z_{\rm map}$, as the fiducial photometric redshift estimate; the percentile-based quantities $z_{16}$ and $z_{84}$ are used to characterize the redshift uncertainty.

We identified M1149-BSG-z5 through our visual inspection of the processed JWST imaging data and catalog.
Fig.~\ref{img} shows the stamps of HST and JWST images of M1149-BSG-z5 at 0.4--5.0\,\micron\ and its composite RGB image.
We interpret M1149-BSG-z5 as an early barred spiral galaxy (with a clump on its southeastern spiral arm) from its complex morphology.
The photometric redshift is inferred as $z_\mathrm{phot} = 5.30 \pm 0.05$, which is tightly constrained by its Lyman break (F435W--F606W), Balmer break (F200W--F277W), and H$\alpha$ excess in the F410M band.
We also cross-matched with the photometric catalog from the CANUCS/Technicolor data release \citep{Sarrouh2026} and the matched ID is CANUCS-5212433 \citep[see also][]{Withers2026}.

% The best-fit SED and redshift probability distribution are shown in the bottom right panel in Fig.~\ref{img}.
% The redshift probability distribution shows a clear single peak at z of 5.30 (maximum probability with minimum $\chi^2$) with 68\% uncertainties of 0.04 and 95\% uncertainty of 0.12. % with... no oversampling... 
% The photometric redshift of M1149-BSG-z5 is tightly constrained by its Lyman-$\alpha$ break (detected as flux jump from HST F435W to JWST F070W), Balmer break (shown as flux increment from F200W to F277W), and H$\alpha$ excess in the F410M band.

\subsection{NIRSpec data}
\label{sec_02c:nirspec}

M1149-BSG-z5 was also observed by JWST NIRSpec micro-shutter assembly (MSA) \citep{jakobsen22} through the G140M/F100LP and G395M/F290LP grating/filter pair (0.9--1.9, 2.8--5.3\,\micron; spectral resolution $R \gtrsim 1000$).
The observation was conducted through Cycle-3 GTO-4527 program (PI: Willott) with a total on-source integration time of 1.7\,hr.
%\xwang{FIXME: We directly used the public, calibrated spectral products from the Mikulski Archive for Space Telescopes (MAST).}
We first used the STScI \texttt{jwst} stage 1 pipeline to process the data, with custom snowball and $1/f$ noise corrections. We then ran \texttt{jwst} stage 2 pipeline up to the photometric calibration stage, the utilized \texttt{grizli} and \texttt{msaexp} \citep{Brammer_msaexp_NIRSpec_analyis_2022} to perform the subsequent steps.

The NIRSpec MSA covers the central bright structure and part of the elongated bar-like feature (shown by the F444W image in Fig.~\ref{img}).
% The 1D spectrum shows a single emission-line redshift, with no evidence for an additional line system at a distinct redshift within the slit. 
We measured a spectroscopic redshift  $z_\mathrm{spec} = 5.102$ from the robust detection of \hbeta, \oiii, \halpha, \nii\ and \sii\ lines.
The spectroscopic redshift is only slightly lower than the $z_\mathrm{phot}$ estimate by $\sim4$\%.
The 1D spectrum and the spectral regions around \hbeta, \oiii, and \halpha\ are shown in the middle and bottom panels of Fig.~\ref{img}, with additional emission-line regions presented in Appendix~\ref{emlines}.

% \xwang{The NIRSpec MSA slitlet covers the central bright structure and part of the bar-like structure (Fig.~\ref{img}). No additional emission lines at different redshifts are detected in the stacked spectrum, confirming that these substructures belong to the same galaxy system.
% Although the NIRSpec MSA slitlet does not directly cover the southern clump, the photometric redshift of this clump ($z_\mathrm{phot}=5.23$) is consistent with that of M1149-BSG-z5, 
% suggesting that it is more likely associated with the same system than being a lower-redshift interloper.}

% Detailed modeling of the spectral properties of M1149-BSG-z5 is beyond the scope of this morphological study.

\begin{figure*}[!t]
    \centering
    \includegraphics[width = \textwidth]{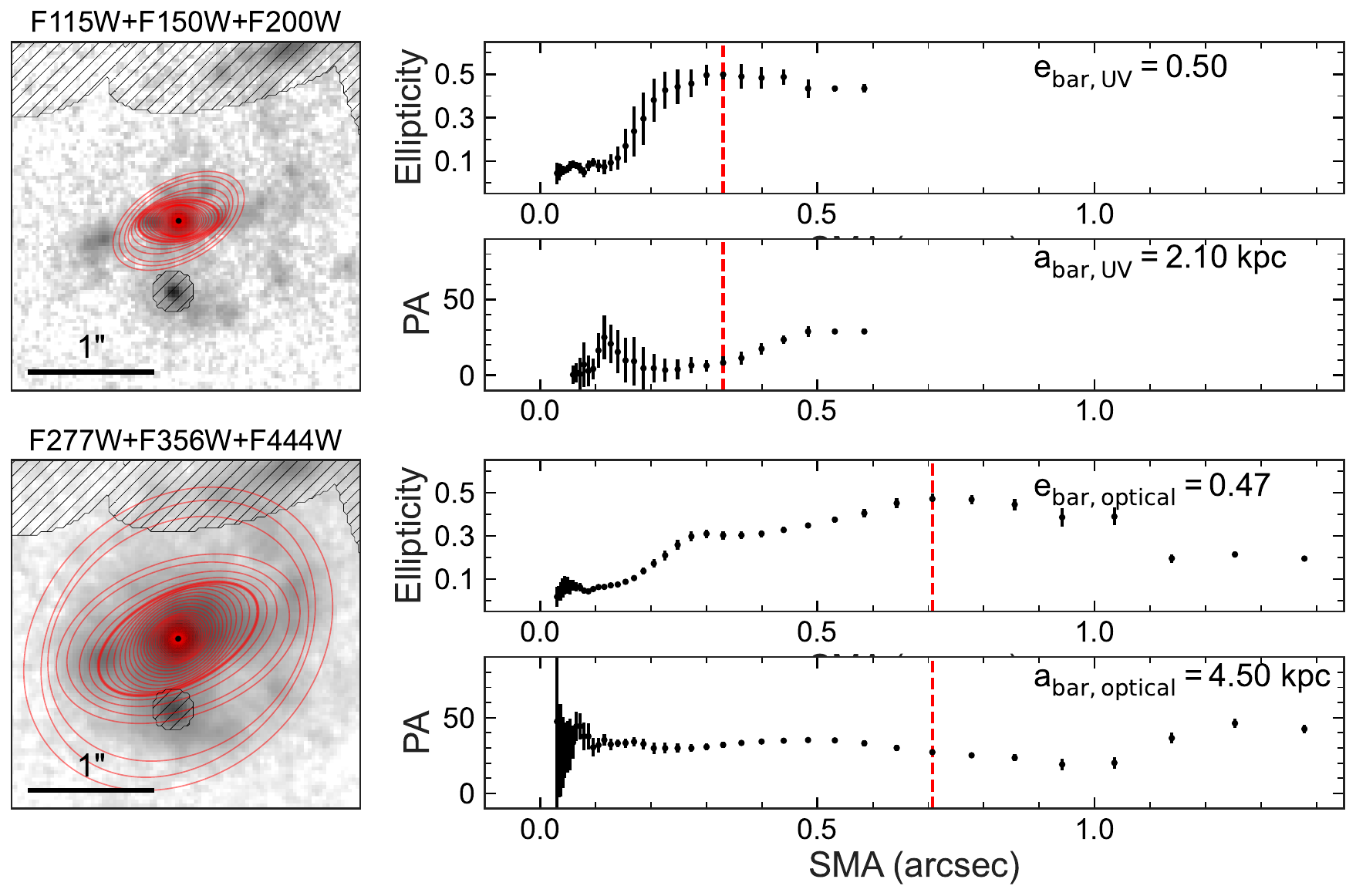}
    \caption{Ellipse isophote fitting and radial profiles of ellipticity and position angle for ellipse isophotes for JWST stacked images of F115W+F150W+F200W, and F277W+F356W+F444W. The southern clump and nearby galaxies are masked in ellipse fitting (black hatches). 
    For short wavelengths (rest-frame UV), the ellipticity reaches maximum $e_{\rm max} \sim 0.50$ at $r \sim 0\farcs33 = 2.10$\,kpc.
    For longer wavelengths (rest-frame optical), the ellipticity reaches maximum $e_{\rm max} \sim 0.47$ at $r \sim 0\farcs71 = 4.50$\,kpc. Bar length $a_{\rm bar}$ is marked as the bold red ellipse in the left panel images and the red dashed line in the radial profiles.}
    \label{isophote}
\end{figure*}

\section{Analysis and Results}
\label{Results}

\subsection{Bar identification}

M1149-BSG-z5 shows a bar-like morphology across multiple filters and in the composite image.
This morphology is not affected by strong gravitational lensing, as the source is located far from the MACS\,J1149 cluster center ($6\farcm8$), where the lensing magnification is expected to be negligible. 
The bar-like structure is clearly present in shorter wavelengths 
%, especially from JWST F090W to F277W 
owing to their better angular resolution, % and signal-to-noise ratios, 
and becomes less distinct at longer wavelengths due to smearing with the redder stellar component of the host galaxy (Fig.~\ref{img}).

A widely adopted method to identify bars is ellipse isophote fitting for galaxy surface brightness.
% As bar structures have typically higher ellipticity and are usually misaligned with the major axis of the galaxy disk, ellipticities of isophotes are expected to increase with radii first and drop at bar typical lengths, while position angles (PAs) keep almost constant in bar regions with a jump change in the disk-dominated regions.
Criteria of bar detection require the ellipticity to rise smoothly to maximum value larger than 0.25, with almost constant position angles (PAs); in the region dominated by the outer disk, the ellipticity would drop by at least 0.1 with PA changing by at least 10\arcdeg\ \citep{Jogee2004, Marinova2007, Guo2023}.
This method is widely used for bar detection in the early Universe \citep[e.g.,][]{Guo2023, Guo2025} and remains robust for high-redshift bars whose intrinsic bar lengths are larger than $2\times$ FWHM (Full Width at Half Maximum; \citealt{Liang2024}).
We performed ellipse isophote fitting with \texttt{PHOTUTILS}
% \footnote{https://photutils.readthedocs.io/} 
\citep{Bradley2025} for stacked JWST images, F115W+F150W+F200W and F277W+F356W+F444W, corresponding to rest-frame UV and optical wavelengths.
The southern clump and nearby galaxies are masked during the fitting. 
We first fit with free parameters to identify the centroid, and then fix the centroid to derive the radial profiles of ellipticities and PAs. 

The ellipse isophotes and radial profiles of ellipticities and position angles for the stacked images are shown in Fig.~\ref{isophote}.
Both rest-frame UV and optical fit are consistent with the standard bar-identification criteria.
% We find that both UV and optical ellipse fitting satisfy bar detection.
In the rest-frame UV stack,
% with better resolution, 
the ellipticity reaches maximum $e_{\rm max} \approx 0.51 $ at $r\approx 0\farcs3$, with almost constant PA. 
% Due to the lower surface brightness at the rest-frame UV bands, the outer disks are not well captured by the ellipse fitting. Even so, 
Although the lower surface brightness in the UV limits the constraints on the outer disk, we can see a mild decrease of ellipticity with PA increment of $\sim 20\arcdeg$ at $r > 0\farcs5$. 
% , and then drops to $\sim 0.4$ with PA increasing. 
This radial change likely captures the inner high surface brightness region of the bar, while the outer bar region and the disk are relatively faint and provide limited constraints on the outer isophotes.
In the rest-frame optical stack, where the outer disks are better captured by the ellipse, the profiles more clearly satisfy the bar detection criteria: the ellipticity rises to maximum $e_{\rm bar}= 0.47$ at $r \sim 0\farcs71$, with almost constant PAs, and drops to $\sim 0.2$ with PA increment of $\sim 20\arcdeg$ at $r > 1\farcs1$.

We estimate the bar typical length $a_{\rm bar}$,% based on ellipse isophote fitting. Bar length $a_{\rm bar}$ 
which is defined as the semi-major axis radius where the bar ellipticity first reaches a maximum value \citep{Athanassoula2002, Jogee2004, Marinova2007, MK2007, Guo2023}.
The ellipse fitting gives
% For M1149-BSG-z5, we measured  
$a_{\rm bar} = 2.10$\,kpc ($0\farcs33$) in the rest-frame UV stack and 4.50\,kpc ($0\farcs71$) in the rest-frame optical stack; both are resolved at $>2 \times$ FWHM of the point-spread function (PSF).
We note that the UV emission primarily traces young, clumpy star-forming regions that are less obscured by dust. 
Therefore, the size measured in the rest-frame optical is more representative of the underlying stellar bar than that in the UV \citep[e.g.,][]{windhorst02}.

We find that the bar length of M1149-BSG-z5 is surprisingly large.
Previous surveys of barred galaxies at $2< z < 4 $ suggest that $a_\mathrm{bar}$ decrease with redshifts \citep[e.g.,][]{Guo2025}. 
However, M1149-BSG-z5 has comparable $a_\mathrm{bar}$ even with those at $z < 1$,
% Moreover, M1149-BSG-z5 remains 
larger than typical barred galaxies at $z \sim 4 $ in both the F200W and F444W measurements \citep{Guo2025, LeConte2026}.
% although the absolute value of $a_{\rm bar}$ may still be affected by clumpy morphology, PSF smearing, the outer disk, and nearby-source treatment.
The elongated bar size suggests that M1149-BSG-z5 is unusually mature in structures for its early cosmic epoch.
However, we caution the possible smearing by the outer disk and the PSF for $a_{\rm bar}$ measurements and further interpretations.

% F200W, ebar = 0.51, abar = 0.33'' = 2.06kpc
% F277W, ebar = 0.55, abar = 0.94'' = 5.87kpc
% F356W, ebar = 0.47, abar = 0.71'' = 4.41kpc
% F444W, ebar = 0.49, abar = 0.86'' = 5.34kpc

% For shorter wavelength (F200W) with better resolution, the ellipticity reaches maximum $\epsilon_{\rm max} \approx 0.5 $ at $r\approx 0\farcs3$, with almost constant position angle, and then drops to $\sim 0.4$ with PA change of $\sim 20 ^\circ$. This radial change possibly captures the inner high surface brightness region of the bar, while the outer bar region and the disk are relatively faint and provide limited constraints on the outer isophotes.
% For F277W and F356W, the ellipticity reaches maximum $\epsilon_{\rm max} \approx 0.45 $ at $r\approx 0\farcs75$ and then drops to $\sim 0.2$ with PA change of $\sim 30^\circ$. The lower ellipticity peak is mainly due to the PSF smearing. The larger radii of ellipticity peak may reflect the detection of the bar edge with better signal-to-noise, or due to the smearing with the spiral arm structure.
% For F444W, ellipticity reaches similar maximum at a similar radius, but declines to $\sim 0.35$ before dropping to $\sim 0.2$ at larger radii.
% This is primarily due to the larger PSF and stronger smearing of the bar and disk regions. 
% Collectively, the isophote fitting further confirms the bar structure in M1149-BSG-z5.

\subsection{Structural analysis: bulge, disk, spiral}
\label{structural_analysis}
\begin{figure*}
    \centering
    \includegraphics[width = \textwidth]{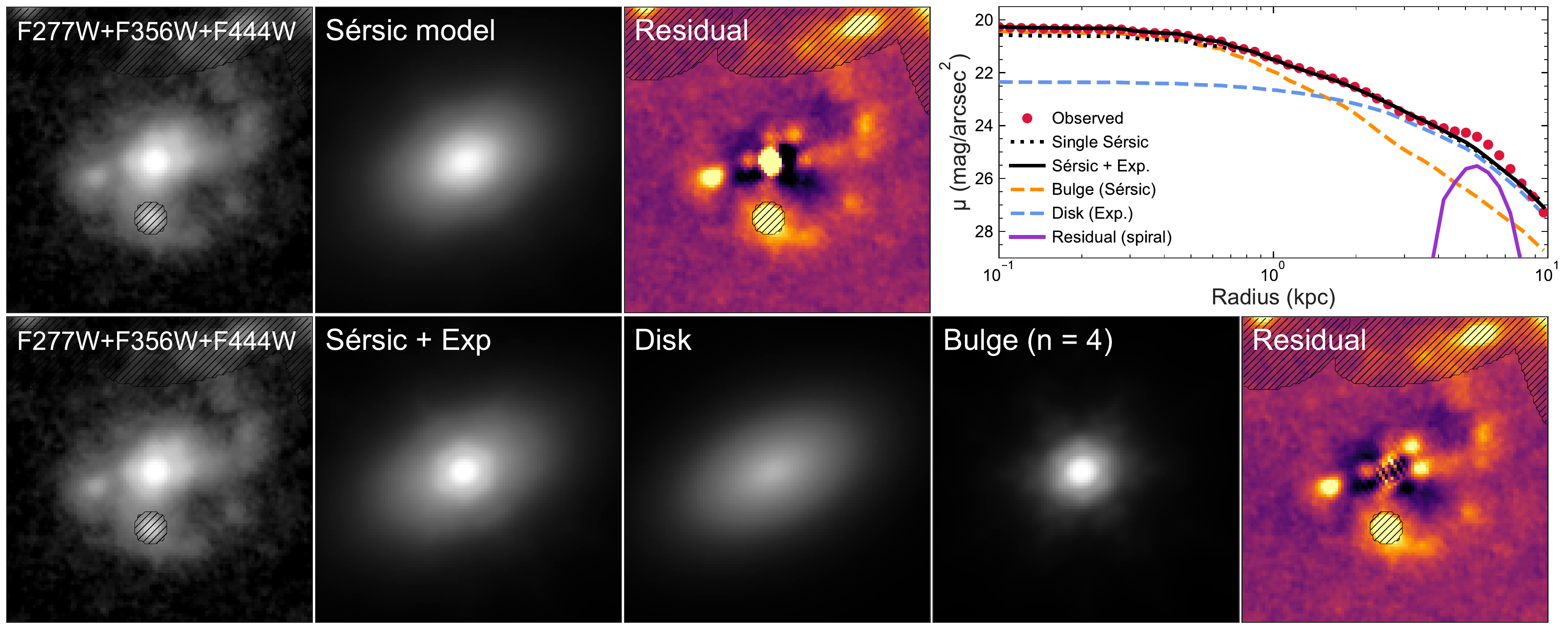}
    \caption{Structural modeling of M1149-BSG-z5 stacked image, F277W+F356W+F444W.
    Top: best-fit model and residual for single Sérsic fit.
    Bottom: best-fit model, sub-component models and residual for bulge + exponential disk fit. The upper-right panel shows the observed and model surface-brightness profiles.
    The single-Sérsic model (black dotted line) underestimates the central brightness, as seen in the residuals, while the bulge+disk model (black solid line) better matches the observed data (orange dots), reducing central residuals and reproducing the concentrated light, suggesting a prominent bulge component.
    Both the bar-like and spiral-arm structures can be seen in residuals.
    }
    \label{struc_fitting}
\end{figure*}

% stellar mass and mean star formation rate in the last 100Myrs of the Bayesian posterior distribution%  in log space
% , and uncertainties by the 16th–84th percentile range.

We quantify the structural properties of M1149-BSG-z5 by fitting the stacked F277W+F356W+F444W image, which traces the rest-frame optical light, using \texttt{PySersic}
% \footnote{\url{https://github.com/pysersic/pysersic}}
\citep{PashaMiller2023}.
We consider both a single-Sérsic model and a Sérsic bulge + exponential disk model.
The PSF is constructed with \texttt{STPSF} \citep{Perrin14}, and the southern clump and nearby galaxies are masked during the fitting.
For the bulge\,$+$\,disk fitting, we fix the bulge Sérsic index to $n=4$.
This choice is motivated by tests in which allowing the bulge Sérsic index to vary yielded poorly constrained values, likely due to the limited spatial resolution, while a pure PSF component provided a worse fit than an extended Sérsic bulge.
We therefore adopt an $n=4$ bulge as a simple empirical representation of the compact central component, rather than over-interpreting the exact Sérsic index value. %as a robustly measured quantity.

The best-fit models and residuals, as well as the 1D surface brightness profiles of the different models and sub-components, are shown in Fig.~\ref{struc_fitting}.
The single Sérsic fitting derives a Sérsic index of $n = 2.37\pm 0.03$ and an effective radius $R_e = 2.61\pm0.02$\,kpc. The bulge\,$+$\,disk fitting derives a disk effective radius of $3.96\pm 0.04$\,kpc and a bulge-to-total light ratio of $B/T = 0.42\pm0.01$.
Compared with the single-Sérsic model, the bulge+disk model better reproduces the central surface brightness and leaves smaller central residuals (see residual maps in Fig.~\ref{struc_fitting}), indicating that an additional compact central component is required to describe the rest-frame optical light distribution.
As the galaxy hosts a prominent bar-like structure, the inferred $B/T$ may be affected by bar light being absorbed into the compact component. Our additional tests with a free bulge Sérsic index give broadly consistent structural parameters, suggesting that this effect is unlikely to qualitatively change our interpretation.
% Compared with single Sérsic fitting, bulge + disk model better reproduces the central surface brightness.
% \xwang{This is evident from the residual map, where the single-Sérsic model underestimates the central brightness, and is further supported by the bulge-to-total ratio of $0.42\pm0.01$, confirming the presence of a strong concentrated bulge component.}

% The structure fitting further emphasizes the existence of bar; 
The residual maps after subtracting these smooth models further highlight the non-axisymmetric structures in M1149-BSG-z5.
Both residual maps show elongated structures. 
Even more strikingly, the residuals show extended spiral-like structure connected to the edge of the bar. Even though the spiral structure is not as grand-design as local spiral galaxies, it shows smooth, continuous distribution instead of combinations of isolated clumps.
We further analyze the surface brightness of the bulge+exp residual, shown as the purple curve in the top right panel of Fig.~\ref{struc_fitting}.
The surface brightness profile of the residual appears a bump that reaches the maximum brightness at 5.52 kpc.
This identified structure, accompanied with the bar and bulge components, suggests M1149-BSG-z5 to be a barred spiral galaxy. 
Even though it may differ from typical mature barred spiral galaxies, it remains interesting to investigate the formation mechanisms of these structures in such an early Universe.

% Both residual maps show elongated structure and extended spiral like structures. The bar structure in the residual maps is not quite clear; however, we note that the bar component was not explicitly included in our structural decomposition. Therefore, the symmetric models (bulge and disk) may partially account for the bar's distribution and overpredict the central light. The dipolar structure in the center regions further confirms the bar feature present in M1149-BSG-z5. 

% Analysis on bar length?
% 1. z = 3: N = 0.65, Rbar sim 3.3kpc
% 2. z = 4.4  no fitting
% 3. z = 4.26, N = 0.9 ± 0.1 ReF444W sim 3.8 ± 0.4 kpc
% 4. z = 2.76, 
% 5. z = 3.8 For the stellar distribution, the best-fitting values of the Sersic ´
% index and effective radius are n = 1.32 ± 0.04 and re = 0.72 ± 0.01
% kpc for the central component of the double Sersic model, and
% n = 0.25 ± 0.01 and re = 1.96 ± 0.03 kpc for the extended component. 

\begin{figure*}[!t]
    \centering
    \begin{subfigure}{\columnwidth}
        \includegraphics[width = \columnwidth]{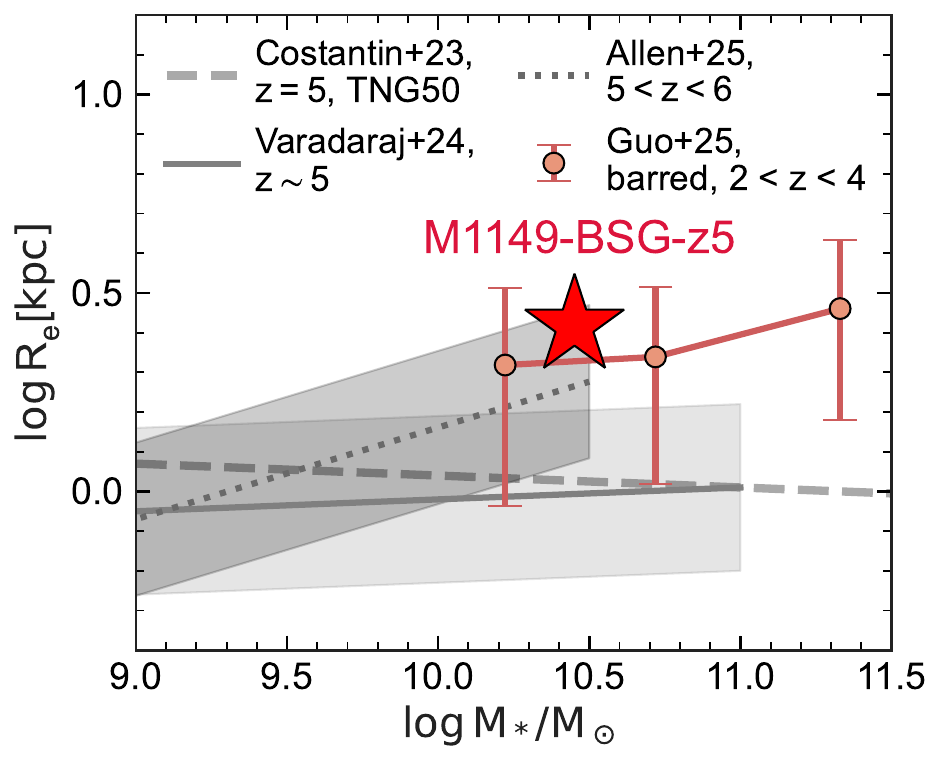}
    \end{subfigure}
    \hfill
    \begin{subfigure}{0.99\columnwidth}
        \includegraphics[width = \columnwidth]{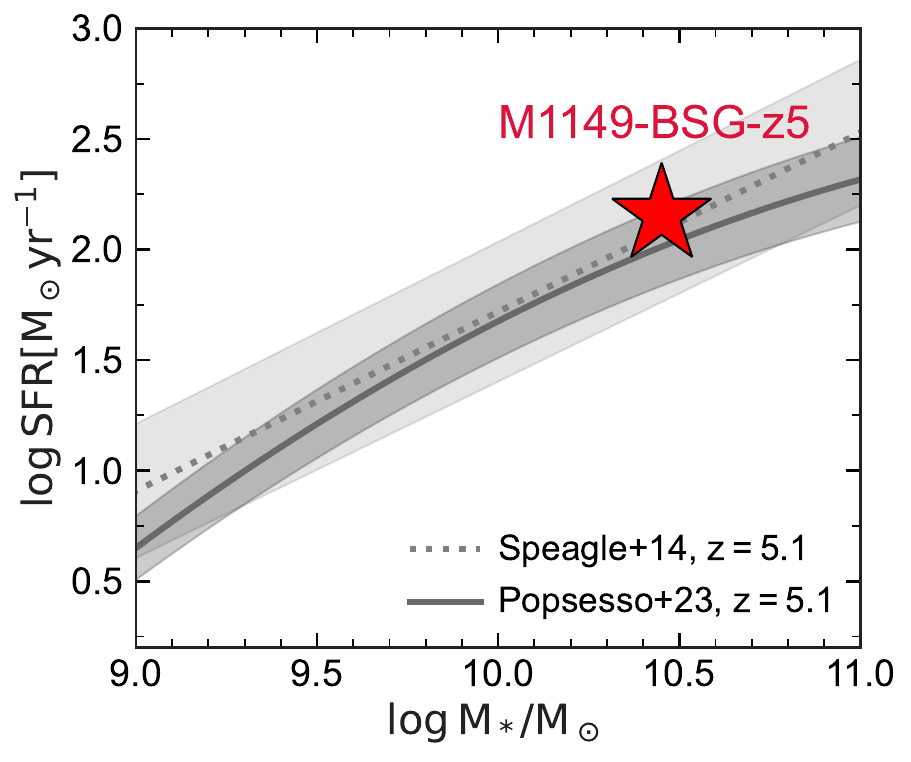}
    \end{subfigure}
    \caption{Left: Location of M1149-BSG-z5 on the size-mass plane. The size-mass distribution of galaxies at similar redshifts, measured with F356W for TNG50 \citep{Costantin2023} and observations \citep{V2024, Allen2025} are included for reference. The size-mass for barred galaxies at $ 2 < z < 4$ \citep{Guo2025} is also shown.
    M1149-BSG-z5 shows relatively larger galaxy size than galaxies at similar redshifts and comparable to barred galaxies at $2 < z < 4$.
    Right: Location of M1149-BSG-z5 on the star-forming main sequence. Parameterized SFMS from \citet{Speagle2014} and \citet{Popesso2023} at $z = 5.1$ are included for reference.
    M1149-BSG-z5 locates on the star-forming main sequence.
    }
    \label{sizemass_SFMS}
\end{figure*}

% \subsection{Location on size-mass plane and star formation main sequence}
\subsection{SED fitting: locations on size-mass and SFMS}
% \begin{figure*}
%     \centering
%     \includegraphics[width = \textwidth]{figures/SED_maps.pdf}
%     \caption{2D maps of star formation rate density, stellar mass surface density, mass-weighted age and dust attenuation ($A_V$) from \texttt{pixedfit} SED fitting with \texttt{pixedfitbin} binning. The stellar population properties are asymmetric. East to the central bulge shows relatively older stellar population age, less star formation rates and less dust, while the west to the bulge shows younger stellar population age with higher star formation rates and more dust. We interpret that this is related to the asymmetric gas inflow from the west of the bar.}
%     \label{SED_maps}
% \end{figure*}

The stellar mass and star formation rate of M1149-BSG-z5 are derived through SED fitting with CIGALE v2025.0 \citep{Burgarella2005, Noll2009, Boquien2019}, using \verb|KRON_S| photometry \citep{Kron1980} with a Kron factor of 1.2, corrected for point-source aperture correction.
% We utilized imaging data from HST F814W, F606W, F435W, JWST F070W, F090W, F070W, F090W, F115W, F200W, F277W, F356W, F444W, F140M, F182M, F210M, F300M, F335M, F410M, F430M, F460M and F480M.
A delayed-$\tau$ star formation history is assumed, together with a \citet{Chabrier2003} initial mass function, \citet{Bruzual2003} stellar population synthesis models, and the \citet{Calzetti2000} dust attenuation law. Both stellar and nebular metallicities are fixed to solar metallicity ($Z = 0.02$).% ($Z_{\odot}$).
The input parameters in the SED fitting are summarized in Appendix \ref{SED_table}.
We adopt the 50th percentile values of the Bayesian posterior distributions%  in log space
, and uncertainties by the 16th–84th percentile range. The star formation rate (SFR) is the mean SFR in the last 100 Myrs.

We adopt the effective radius derived from the single-Sérsic fitting ($R_e = 2.61$\,kpc) as the characteristic global size.
The location of M1149-BSG-z5 on the size-mass plane is shown in the left panel of Fig.~\ref{sizemass_SFMS}.
Median trends of size-mass for observed and simulated galaxies for rest-frame optical filters (F356W) at similar redshifts are included for reference.
M1149-BSG-z5 is typically larger than observed or simulated galaxies at similar redshifts \citep{Costantin2023, V2024, Allen2025}.
We also include the median size-mass for barred galaxies at $2 < z < 4$ from the JWST CEERS survey \citep{Guo2025}.
We notice that M1149-BSG-z5 is comparable in size with barred galaxies at $2 < z < 4$ and slightly higher than the average trend.
% The measurement for CEERS is from F356W for galaxies with lower redshifts and therefore longer wavelength in rest-frame compared with M1149-BSG-z5.
% Therefore, the effective radius of M1149-BSG-z5 at similar wavelength could be even larger than that measured with F356W.
The large size of M1149-BSG-z5 and its high stellar mass ($10^{10.45}  \rm M_\odot$) suggest that M1149-BSG-z5 may represent an early-assembled, evolved extended disk.
The locations of M1149-BSG-z5 on the SFR vs. stellar mass plane are shown in the right panel of Fig.~\ref{sizemass_SFMS}.
% Parameterized SFMS from \citet{Speagle2014} and \citet{Popesso2023} are included as general benchmarks.
% and data from observations are also included... 
M1149-BSG-z5 lies on the star-forming main sequence at $z \sim 5$.

\subsection{Emission-line analysis: AGN and metallicity}

\begin{figure*}
    \centering
    \begin{subfigure}{0.985\columnwidth}
        \includegraphics[width = \columnwidth]{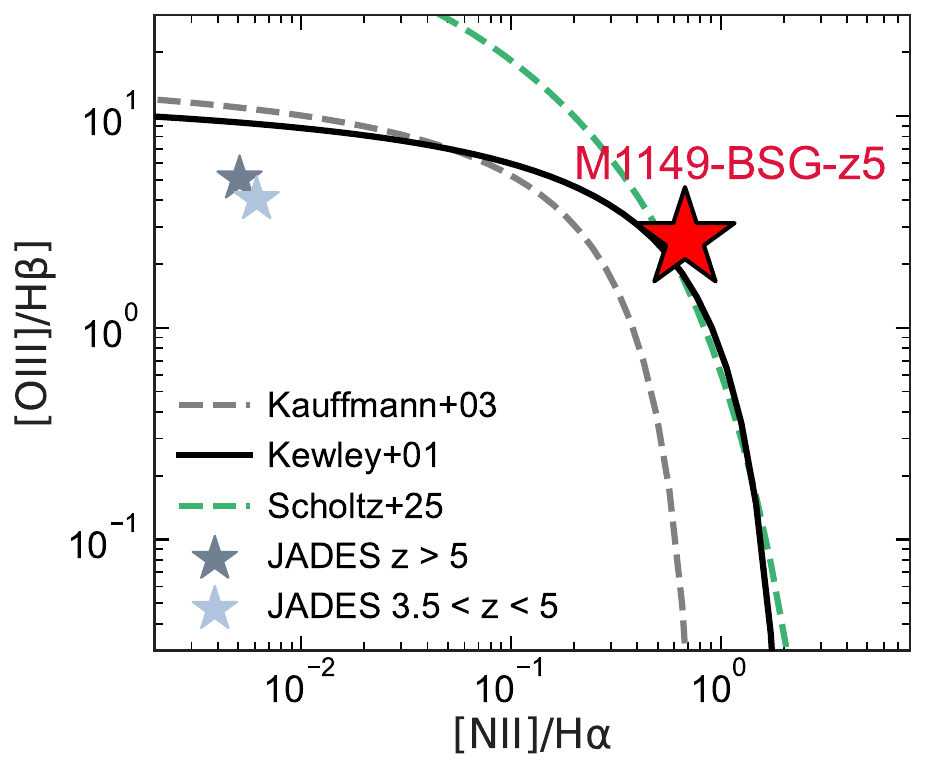}
    \end{subfigure}
    \hfill
    \begin{subfigure}{1.01\columnwidth}
        \includegraphics[width = \columnwidth]{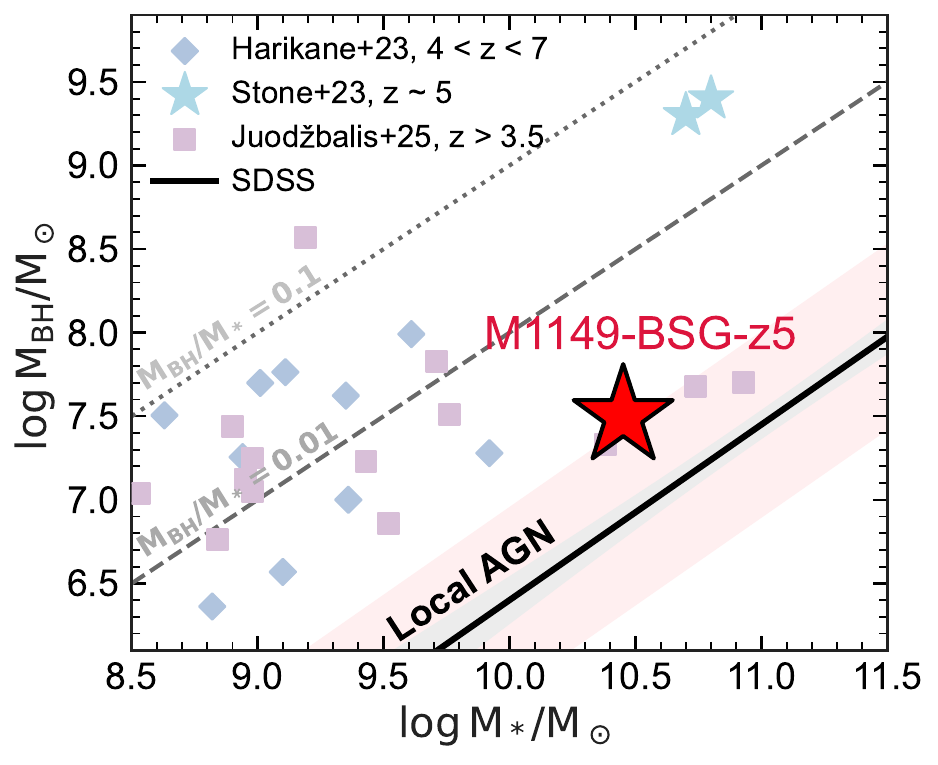}
    \end{subfigure}
    \caption{Left: Location of M1149-BSG-z5 on the N2-BPT diagram. The dashed gray curve shows the AGN classification from \citet{Kauffmann2003}, and the black solid curve shows the theoretical maximum-starburst boundary of \citet{Kewley2001}. The green dashed line shows the classification of high redshift galaxies from \citet{Scholtz2025}. The line ratios measured from stacked spectra of JADES AGNs at $3.5 < z < 5$ and $z > 5$ \citep{J2026} are shown as gray stars for reference.
    Right: Location of M1149-BSG-z5 on the $M_{\rm BH} - M_*$ relation. The black solid line and gray shaded region show the best-fit relation with $1\sigma$ uncertainty for local AGNs from SDSS \citep{Reines2015}, and the light red shaded region shows the rms scatter. Broad-line AGN samples (including quasars and little red dots) at similar redshifts are shown for comparison \citep{Harikane2023, Stone2024, J2026}.}
    \label{bpt_bh}
\end{figure*}

The NIRSpec spectrum has clear detections of [O\,{\sc i}], [O\,{\sc ii}], H$\beta$, [O\,{\sc iii}]4959,5007, H$\alpha$, [N\,{\sc ii}]6548,6584 and [S\,{\sc ii}] (Fig.~\ref{img}, Appendix~\ref{emlines}).
We therefore diagnose the ISM and ionizing source properties in M1149-BSG-z5.
We model each emission line with a Gaussian profile to measure the line flux and line width. The continuum is modeled by a constant flux offset around each line, and the fit is performed at $\pm700$ \AA\ of each line.
The best-fit parameters are determined by the Levenberg-Marquardt non-linear least-squares algorithm. %  implemented in the \texttt{SciPy} module.
We fix the [O\,{\sc iii}]4959/5007 ([N\,{\sc ii}]6548/6584) doublet ratio to 2.98 (3.00) and only fit the stronger line flux as a free parameter.
In addition to these, the H$\alpha$ line profile is not well fit with a single Gaussian, so a two-component (narrow+broad) Gaussian profile is adopted for the H$\alpha$ line ($\Delta \rm BIC = BIC(narrow + broad) - BIC(narrow\,only) \approx -67$). The best-fit models are shown in Fig.~\ref{img} and Appendix~\ref{emlines}. 
Appendix~\ref{emlines} presents a table of the best-fit line fluxes.

% The H$\alpha$/H$\beta$ ratio clearly deviates from the theoretical prediction under the case B recombination ($3.7\pm0.6$ for the narrow line), suggestive of the dust attenuation.
% Assuming \citet{Calzetti2000} attenuation law with $R_V=4.05$, we obtain $A_V=0.9\pm0.5$ mag attenuation.
% In the following, we correct for the dust attenuation to compute the line ratios and associated physical properties.

M1149-BSG-z5 shows an extra broad component in $\rm H\alpha$ with FWHM of $\sim 1900\,\rm km\,s^{-1}$, suggesting the presence of a broad-line AGN.
% \citep{Ubler2023, Harikane2023, Kocevski2023, Greene2024}.
We investigate the location of M1149-BSG-z5 on the N2-BPT diagram ([O\,{\sc iii}]$/\rm H\beta$ - [N\,{\sc ii}]/$\rm H\alpha$; \citealt{Baldwin1981}) with flux ratios of the narrow components, as shown as the left panel of Fig.~\ref{bpt_bh}.
M1149-BSG-z5 lies above both the empirical 
star-forming region of \citet{Kauffmann2003} and the theoretical maximum-starburst boundary of \citet{Kewley2001}. 
M1149-BSG-z5 resides in the region dominated by low-redshift AGNs, which show higher [N\,{\sc ii}]$\rm /H\alpha$ with harder ionization conditions.
In contrast, the vast majority of JWST-discovered high-redshift AGNs show systematically offsets with higher ionization and lower metallicity, overlapping with the star-forming region in the N2-BPT diagram \citep[e.g.,][]{Nakajima2022, Ubler2023, J2026}.

% The detections of relatively strong [N{\sc ii}]6584, [O{\sc i}]6302, and [S{\sc ii}]6716,6731 lines, in addition to the broad-line H$\alpha$ component, implies the presence of AGN as the predominant ionizing source of the BSG-z5 spectrum.
We estimate the black hole mass based on the extinction-corrected luminosity and the line width of the broad H$\alpha$ component.
The extinction is corrected using Balmer decrement assuming \citet{Calzetti2000} attenuation law with $R_V=4.05$.
Based on the empirical calibration by \citet{Reines2013ApJ}, we derive the BH mass as $M_{\rm BH}=10^{7.5\pm0.2}\ M_\odot$, which corresponds to the BH-to-stellar mass ratio of $M_{\rm BH}/M_\star \sim 10^{-3.0}$ in M1149-BSG-z5 (right panel of Fig.~\ref{bpt_bh}).
This ratio is relatively lower than the majority of the AGNs at $3.5 < z< 7$ discovered through JWST surveys \citep[e.g.,][]{Harikane2023, Stone2024, J2026}, but comparable to broad-line AGNs in the local Universe \citep{Reines2015}.
This places M1149-BSG-z5 near the local AGN locus in the $M_{\rm BH}$--$M_\ast$ plane, rather than among the strongly overmassive black holes commonly reported at high redshift. M1149-BSG-z5 may therefore represent a more mature and evolved AGN host system already in place at $z\sim5$.

We finally analyze the metal properties based on the narrow-line flux ratios.
The oxygen abundance measured with the $\hat{R}$ indicator ($=0.47 \log_{10}({\rm O2}) + 0.88\log_{10}({\rm O3})$; \citealt{Laseter2024AAP}) under the calibration by \citet{Sanders2026ApJ} gives $12+\log({\rm O/H})=8.43^{+0.08}_{-0.12}$ ($\sim50\%$ solar), suggesting that M1149-BSG-z5 is a metal-enriched system.
This places M1149-BSG-z5 at the high-mass, high-metallicity end of the mass-metallicity relation at this redshift \citep{Faisst2026}.
This estimate should be treated with caution, as AGN ionization conditions are complex and standard star-forming metallicity calibrations may not be directly applicable. Nevertheless, its BPT location is close to that of low-redshift AGN rather than the low-[N\,{\sc ii}]$\rm /H\alpha$, high-[O\,{\sc iii}]$\rm /H\beta$ regime often associated with low-metallicity high-redshift systems \citep[e.g.,][]{Ubler2023}. Together with the elevated [N\,{\sc ii}]/$\rm H\alpha$ ratio, the emission-line ratios are difficult to reconcile with a very metal-poor interpretation, and instead suggest that the gas is already chemically enriched in M1149-BSG-z5.
Taken together, %the BPT location of M1149-BSG-z5, its local-like $M_{\rm BH}/M_\ast$ ratio, and its metal-enriched emission-line properties 
the emission-line properties consistently point to a chemically evolved and comparatively mature galaxy already in place at $z\sim5$.

\section{Discussion}
\label{Discussion}
% \subsection{A typical bar?}
\subsection{Early emergence of bar at $z \sim 5$}
% The results presented above show that M1149-BSG-z5 is an unusually complex and evolved system at $z\simeq5.1$.
% The morphological and structural analyses identify a prominent bar-like component embedded in a disk with spiral-like features, while the SED and emission-line analyses indicate a massive, chemically evolved system with nuclear activity.
% M1149-BSG-z5 therefore provides a case in which a bar-like structure is observed in an already evolved galaxy within the first $\sim1$\,Gyr of cosmic history.

The results above show a prominent bar-like structure at $z\simeq5.1$ hosted by an unusually evolved galaxy. As a massive, chemically evolved system with nuclear activity, M1149-BSG-z5 provides a rare case in which a bar-like structure may have already emerged within the first $\sim1$\,Gyr of cosmic history.

A critical concern at $z \sim 5$ is whether the observed feature is a real stellar bar or a group of asymmetric clumps.
Although spatially resolved kinematics are required for dynamical confirmation, the current photometric evidence favors the presence of a bar.
% Without kinematic measurements, we cannot confirm that M1149-BSG-z5 is a real barred spiral galaxy. However, both the ellipse isophote analysis and structure fitting support the existence of bar; t
The ellipse isophote analysis satisfies the standard photometric criteria for bar identification, and the structural fitting supports an elongated central component beyond the smooth bulge+disk model.
Moreover, the bar-like structure is more elongated and horizontal with smooth light distribution instead of isolated clumps, and these features are more conspicuous in the longer wavelength filters where the evolved stellar population dominates the light.
% Secondly, the horizontal elongated structure suggests 
% is more likely to relate to a non-axisymmetric potential.
% Finally, the observed dust attenuation in bar regions are stronger than other regions, and the asymmetric star formation across the bar region is consistent with bar-driven gas inflows and shocks.

As summarized in Sec.~\ref{Intro}, higher bar fractions are revealed by JWST at $z > 1.5$, and bar emergence is dated back to $z \sim 3 - 5$. This challenges the expected epoch of disk settling \citep[e.g.][]{Wisnioski2015, Simons2017, FW2020}.
The discovery of M1149-BSG-z5, a bar candidate at $z = 5.1$ pushes the frontier even further, %, and the stellar population analysis suggests its bar could have been established as early as $z \sim 6$.
and the structural and global properties of M1149-BSG-z5 consistently imply an early-assembled and structurally evolved galaxy.
This suggests that the disk condition required for prominent non-axisymmetric structures can be established much
% dynamically cold disks could assemble and stabilize much 
earlier in the Universe than conventionally expected, and this interpretation is also in line with recent reports of cold gas disks extending to $z \gtrsim 6$ \citep[e.g.,][]{Smit2018, Posses2023, rowland24, Fujimoto2025}.

The physical nature of bars at high redshifts, however, could be quite different from that of bars in the local Universe.
% Different with mature bars in the local Universe, 
The bar-like structure in M1149-BSG-z5 shows asymmetric rest-frame UV brightness (Fig.~\ref{img}) with multiple residual clumps in the disk (Fig.~\ref{struc_fitting}).
% The bar-like structure in M1149-BSG-z5 shows different rest-frame UV brightness on its two sides (Fig.~\ref{img}) with multiple residual clumps in the disk (Fig.~\ref{struc_fitting}).
We apply regional SED fitting to the bulge and the two sides of the bar (Appendix~\ref{app_sed_sub}), finding that the UV-brighter side is slightly younger and less obscured.
These properties may trace recent star formation, dust attenuation, or stellar-population variations in a gas-rich disk, potentially linked to bar-driven gas inflows \citep[e.g.,][]{Pastras2025,Jolly2026, Pastras2026}, or to perturbations and inhomogeneous star formation at this early epoch.
A simple dynamical estimate using the total stellar mass ($10^{10.45}M_\odot$), bar length ($a_{\rm bar}=4.50\,\rm kpc$) gives $V_{\rm c}\sim\sqrt{GM_\star/a_{\rm bar}}\sim164\,\rm km\,s^{-1}$ and an orbital timescale of $T_{\rm orb}\sim2\pi a_{\rm bar}/V_{\rm c}\sim168\,\rm Myr$, comparable to the SED ages. Although highly uncertain, this estimate indicates that the bar- and spiral-like structures may still be in a dynamically young or early assembly stage.
% It is unclear whether the bar structure in M1149-BSG-z5 is long-lived, secularly evolving, or a temporary phase of early structure formation.
High-resolution kinematic observations (e.g., with ALMA or JWST/NIRSpec) are required to analyze its velocity field and reveal the dynamical property of this exceptionally early structure.

% Furthermore, the spiral structure suggest that.... the loose structures.... bar evolution... 

\subsection{Potential bar formation mechanisms}
Formation mechanisms of bars in the early universe have been investigated among different simulations.
Some simulations suggest that bars in the early Universe are mainly induced by interactions or mergers \citep[e.g.][]{Bi2022, Fragkoudi2025}; this is supported by observations that the barred galaxies at $z > 1.5$ are more likely to have companion galaxies \citep{Guo2025}.
Some other simulations, however, suggest that bars emerge from instabilities in early assembled galaxies with specific halo properties \citep{Rosas-Guevara2022}.
Nevertheless, barred galaxies are found to be consistently more baryon dominated among different simulations \citep{Rosas-Guevara2022, Fragkoudi2025}.
This baryon dominance is indicated by models and simulations to sufficiently shorten the bar formation timescale and allow bars to emerge at high redshifts \citep{BH2023, BH2024}.
However, properties of the simulated barred galaxies, including stellar population age and gas fractions, are inconsistent among different simulations \citep{Rosas-Guevara2022, Fragkoudi2025}.

\subsubsection{Environment and tidal interactions}
\label{Env}

\begin{figure*}
    \centering
    \begin{subfigure}{0.55\textwidth}
        \includegraphics[width = \columnwidth]{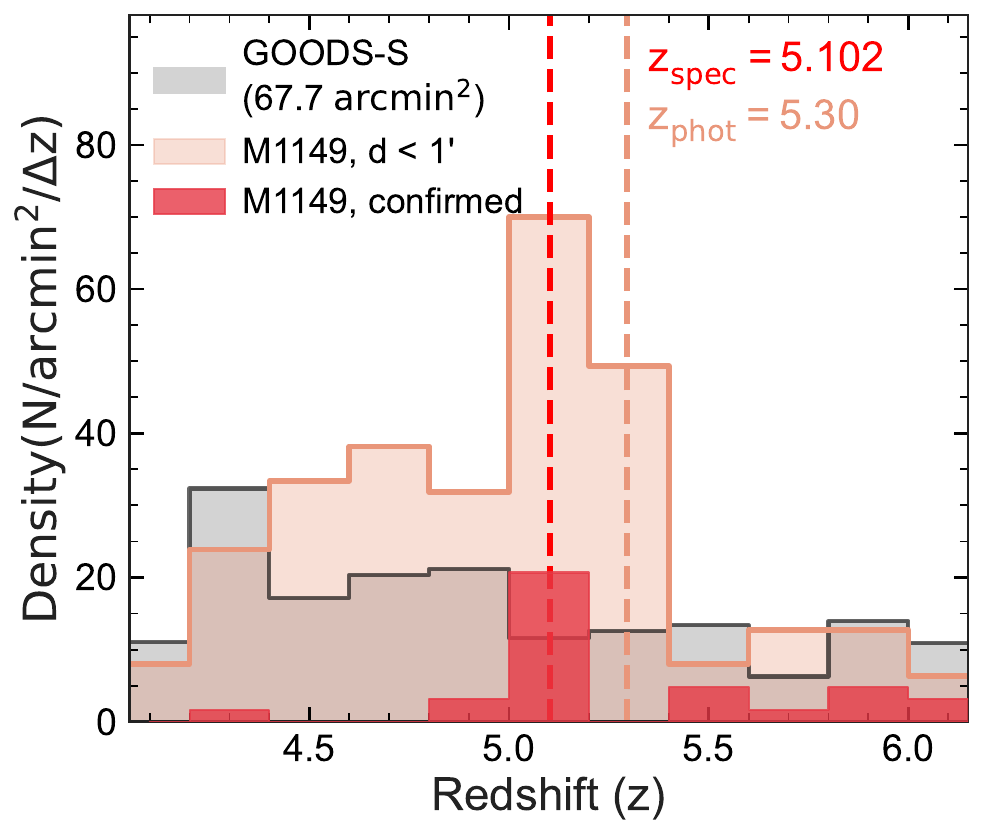}
    \end{subfigure}
    \hfill
    \begin{subfigure}{0.43\textwidth}
        \includegraphics[width = \columnwidth]{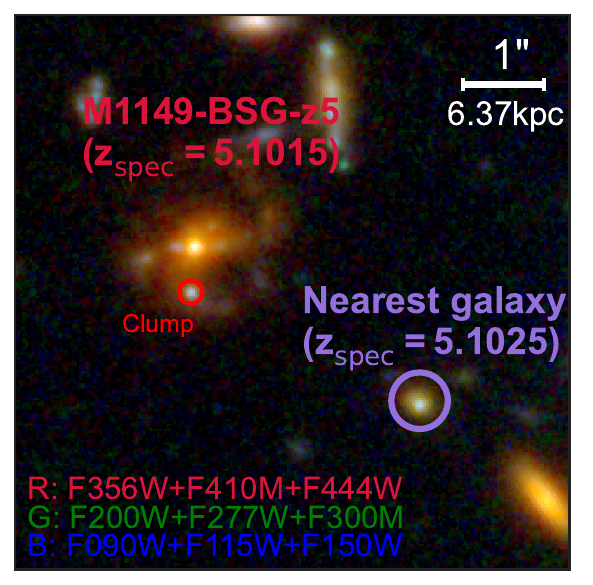}
        \vspace{1.5em}
    \end{subfigure}
    \caption{Left: Number density distribution of redshifts for galaxies within 1 arcmin around M1149-BSG-z5, compared with galaxies in GOODS-South \citep{Hainline2024}. Sources are selected with $\rm m_{F444W} < 29$ and $(z_{84} - z_{16})/(1 + z_{50})< 10\% $.
    Photometric-redshift distribution for galaxies in the M1149 field around M1149-BSG-z5 is shown in shaded pink, with spectroscopic redshift confirmed galaxies shown in red, and the distribution for GOODS-South is shown in shaded gray. 
    The M1149 distribution peaks at $z_{\rm phot}\simeq5.30$, and the spectroscopic sample peaks at $z_{\rm spec}\simeq5.10$, indicating that M1149-BSG-z5 resides in an overdense region.
    Right: Location of a companion candidate of M1149-BSG-z5 with $z_{\rm spec} = 5.1025$, separated from M1149-BSG-z5 by a projected distance of 21.2 kpc.
    }
    \label{pz_env_companion}
\end{figure*}

We investigate the environmental effect on the bar formation of M1149-BSG-z5. 
We start from the photometric catalog (Sec.~\ref{sec_02b:phot}), select galaxies with F444W magnitude $< 29$ for completeness, and require $(z_{84} - z_{16})/(1 + z_{50})< 10\% $ for reliable photometric measurements.
We calculated the number density distribution of photometric redshifts for galaxies within the projected radius of 1\arcmin\ around M1149-BSG-z5. For comparison, we include
% calculate the redshift distribution for 
galaxies in GOODS-S field \citep{Hainline2024} with the same magnitude and photometric redshift selection.
The distributions are shown in the left panel of Fig.~\ref{pz_env_companion}.

The M1149 field shows a pronounced peak at a redshift consistent with M1149-BSG-z5 photometric redshift.
We quantify this enhancement using a local photometric overdensity,
$1+\delta_{\rm phot}=(N_{\rm M1149}/A_{\rm M1149})/(N_{\rm GOODS-S}/A_{\rm GOODS-S}),$
where $N_{\rm M1149}$ and $N_{\rm GOODS-S}$ are measured within $5 < z < 5.4$ to cover the spectroscopic redshift of M1149-BSG-z5.
This gives $1+\delta_{\rm phot}=4.9$, or $\delta_{\rm phot}=3.9$.
We therefore interpret the environment of M1149-BSG-z5 as a moderate local photometric overdensity.
% The enhanced number density, relative to both the GOODS-S field and other redshift intervals in the M1149 field, suggests that M1149-BSG-z5 resides in a local overdensity.

% Compared with the GOODS-S field, 
This interpretation is further strengthened by the spectroscopic redshift distribution.
% Taking the benefits of spectroscopic observations in the same field, we further investigate the $z_{\rm spec}$ distributions of surrounding galaxies. 
We directly adopted the public NIRSpec spectra of Cycle-3 GTO-4527 program (PI: Willott; Sec.~\ref{sec_02c:nirspec}), which provide 168 spectra of galaxies within $\sim 2$ arcmin in the M1149 field, and calculated the redshifts based on [O\,{\sc iii}] and \halpha\ emission lines. 
The $z_{\rm spec}$ distribution is shown as red histograms in Fig.~\ref{pz_env_companion}.
It shows a narrow redshift spike at $z_{\rm spec}\simeq5.10$, with 11 neighboring galaxies spectroscopically confirmed in the interval $5.10<z_{\rm spec}<5.15$ within $1\arcmin$ of M1149-BSG-z5, or 12 galaxies when including M1149-BSG-z5 itself.
As the spectroscopic sample is possibly biased and incomplete, we do not use it to derive an independent spectroscopic overdensity.
Instead, the spectroscopic redshift spike provides an independent confirmation that the photometric overdensity is associated with a real structure at $z\simeq5.1$.
% The $z_{\rm spec}$ confirmed galaxies around M1149-BSG-z5 shows a peak at 5.102, further confirming the overdensity M1149-BSG-z5 lives in.
% Several simulations have suggested that bars and spiral arms in the early universe are mostly induced by interactions or mergers \citep{Bi2022, Fragkoudi2025}.
As overdense environment leads to a higher frequency of interactions and mergers, this may suggest that the bar in M1149-BSG-z5 is merger- or interaction-induced.

Previous JWST observations have found that barred galaxies are more likely to have companion galaxies at $z > 1.5$ \citep{Guo2025}.
% We further checked the projected distances to M1149-BSG-z5 of galaxies with similar redshifts in the same field.
The nearest galaxy to M1149-BSG-z5 is located $21.2$\,kpc away in projected distance, with $z_{\rm spec} = 5.1025$ (shown in the right panel of Fig.~\ref{pz_env_companion}).
% This redshift difference corresponds to $\sim 3400\rm\ km/s$ of $\Delta V$, much larger than the typical velocity difference of galaxy pairs.
Such a small redshift difference ($\Delta v \sim 44$\,$\rm km\,s^{-1}$) implies this galaxy as a companion to M1149-BSG-z5.
SED fitting to this companion candidate with the same SED assumptions as for M1149-BSG-z5 (Table~\ref{SED_para_table}) suggests $\log(M_\star/M_\odot)=8.65\pm0.02$, implying a low stellar-mass ratio of $\sim0.016$.
% Given this low mass ratio, it remains unclear whether tidal perturbations from this companion could have played a significant role in forming or growing the bar-like structure.
In addition, the bright clump to the south of M1149-BSG-z5 may also support an interaction-related origin \citep[e.g.,][]{Wu2023, Wang2025, Xiao2025}, although its nature remains uncertain because no spectroscopic redshift is available and its photometric redshift is $z_{\rm phot}=5.23$.
% However, its nature remains uncertain, as no spectroscopic redshift is available and its photometric redshift, $z_{\rm phot}=5.23$, is offset from that of M1149-BSG-z5.
While the low stellar-mass ratio of the companion suggests that its dynamical impact may be modest, the presence of the close companion and the possible southern clump indicate that M1149-BSG-z5 is not isolated and may have experienced weak external perturbations.
The overdense environment also places M1149-BSG-z5 in line with emerging examples of massive barred disks in protoclusters \citep[e.g.][]{Umehata2025, Boogaard2026}, while isolated barred disks have also been reported \citep{Huang2025}.
% Overall, the close companion candidate and the possible southern clump suggest that M1149-BSG-z5 is not isolated, but the available data do not establish a direct link between external perturbations and its bar- and spiral-like morphology.
Deeper spectroscopic observations and simulations are needed to test this scenario.

\subsubsection{Gravitational instability in baryon-dominated systems}

Another important mechanism of bar formation at high redshift is gravitational instability.
Recent cosmological simulations suggest that high-redshift barred galaxies tend to assemble early and are baryon-dominated \citep{Rosas-Guevara2022, Fragkoudi2025}.
Theoretical frameworks further emphasize that early bar formation requires a high disk (baryon) fraction \citep{BH2023}, while abundant gas can sufficiently accelerate the bar formation timescale \citep{BH2024}.

M1149-BSG-z5 shows several properties consistent with this picture.
Although confirming baryon dominance requires kinematic data, its high Sérsic index, large bulge-to-total ratio, and concentrated central light indicate that a substantial baryonic component has already assembled in the inner region. M1149-BSG-z5 shows active ongoing star formation, and the detection of AGN activity further suggests efficient central fueling and rapid nuclear growth with gas inflow \citep[e.g.][]{Umehata2025, Huang2025}.
Together with its clumpy morphology, these properties point to a baryon-rich and gas-rich system, favorable for rapid bar onset through gravitational instability.
% Although confirming baryon dominance requires kinematic data, our photometric analysis provides some clues.
% First, M1149-BSG-z5 hosts a highly concentrated bulge, which is more likely to form in baryon dominated systems.
% Secondly, M1149-BSG-z5 resides on the star formation main sequence, and the stellar population maps indicate recent gas accretion and inflow. \xwang{(TBD)}
% This suggests that M1149-BSG-z5 is gas rich, where bar formation timescale can be efficiently shortened.
Furthermore, the overdense environment of M1149-BSG-z5 not only indicates higher possibility and frequency of interactions, but also implies early assembly and accelerated star formation \citep{Helton2024, Champagne2025, Galbiati2025, Morishita2025, WangX2025}.
Therefore, M1149-BSG-z5 is likely to be baryon-dominated and gas-rich with early assembly, consistent with rapid bar onset via gravitational instability.

The formation channel of M1149-BSG-z5 bar still requires further observational constraints and more comparisons with simulations.
We also notice that the existence of a bulge may stabilize the disk and weaken the bar structure, and gas inflow driven by bar can further lead to central star formation and weaken bar \citep{BH2024}.
While many detected barred galaxies at $z > 3$ show disky structures with Sérsic $n \sim 1$ \citep[e.g.][]{Costantin2023_bar, Smail2023, Amvrosiadis2025}, M1149-BSG-z5 shows a high Sérsic index, high $B/T$ and AGN activity. We note that the $B/T$ may be affected by unresolved AGN emission or the bar component.
Further observations are needed to investigate the co-existence and co-evolution of bars and bulges at $z \sim 5$.

% \subsection{Possible analogs in local universe}
% Asymmetry?
% More discussion on the asymmetry property... 

\section{Summary}
\label{Conclusions}

With JWST and HST imaging, we report a candidate for one of the earliest known barred spiral galaxies, M1149-BSG-z5 at $z = 5.102$.
The basic properties and analysis of M1149-BSG-z5 are summarized as follows:
\begin{enumerate}
    \item Both isophote ellipse fitting and structural modeling support a presence of bar in M1149-BSG-z5. The bar features in ellipse fitting are present in both rest-frame UV and optical. The structural fitting suggests that M1149-BSG-z5 has extended spiral-arm structures.
    \item M1149-BSG-z5 is a massive star forming galaxy with $M_* = 10^{10.45\pm 0.05}\rm M_\odot$ and SFR = $10^{2.16^{+0.07}_{-0.08}}\rm M_\odot/yr$, residing on star-forming main sequence.
    It hosts a concentrated bulge with global Sérsic $n = 2.37$ and $R_e = 2.61 \rm \ kpc$. Its size is larger than typical galaxies at $z \sim 5$, but comparable to barred galaxy sizes at $2 < z < 4$.
    \item M1149-BSG-z5 hosts an AGN, as indicated by both a broad $\rm H\alpha$ component and its location in the AGN region of the BPT diagram. It has a black-hole-to-stellar mass ratio of $M_{\rm BH}/M_\ast \sim 10^{-3.0}$, lower than those of many high-redshift AGNs and comparable to local AGNs. Together with its BPT location and high metallicity ($\sim50\%$ solar), we conclude that M1149-BSG-z5 is a massive, chemically evolved galaxy at $z\sim5$.
    % \item M1149-BSG-z5 shows asymmetric stellar population across the bar. The left bar region has older stellar population with less star formation and less dust, with a mass-weighted age of $\sim 0.16$ Gyr, suggesting that the stellar bar structure may have started assembling as early as $z \sim 6$. The right bar shows younger stellar populations with higher star formation and more dust. This asymmetry likely reflects recent gas accretion or bar-driven gas inflows.
    \item The photometric redshift distribution of galaxies within 1\arcmin\ of M1149-BSG-z5 shows a peak at $z_{\rm phot} = 5.3$, with spectroscopic redshift distribution peaking at $z_{\rm spec} = 5.1$. This indicates that M1149-BSG-z5 lives in an overdense region. The nearest galaxy to M1149-BSG-z5 with a similar redshift is projected 21.2 kpc away with $z_{\rm spec} = 5.102$, suggesting that interaction may play a role in the bar formation of M1149-BSG-z5.
    
\end{enumerate}

The discovery of M1149-BSG-z5 and its structural and global properties suggests that bars emerge as early as $z > 5$.
While its overdense environment and companion galaxy candidate support the interaction-driven bar formation scenario,
the physical properties, including the light distribution, star formation and early assembly, are broadly consistent with accelerated bar formation via gravitational instabilities in gas-rich baryon-dominated systems. 
Further follow-up observations, particularly kinematic measurements of M1149-BSG-z5, would be the key to confirm its baryon dominance and help to determine the bar formation mechanisms in the early universe.

\begin{acknowledgments}

XW acknowledges further support by Westlake University and the National Science Foundation of China (Grant No. 11821303 to SM).
YA acknowledges support from the Dunlap Institute, funded through an endowment established by the David Dunlap family and the University of Toronto.
RAW acknowledges support from NASA JWST Interdisciplinary Scientist grants NAG5-12460, NNX14AN10G and 80NSSC18K0200 from GSFC.
KK acknowledges the support by JSPS KAKENHI Grant Numbers JP22H04939, JP23K20035, and JP24H00004.
AZ acknowledges support by the Israel Science Foundation Grant No. 864/23.
DE acknowledges support from the Spanish Ministry of Science and Innovation; projects PID2023-150178NB-I00, PID2023-149578NB-I00, PID2020-114414GB-100 and PID2020-113689GB-I00 financed by MCIN/AEI/10.13039/501100011033, and by FEDER, UE; project P20-00334  financed by the Junta de Andaluc\'{i}a; and project A-FQM-510-UGR20 of the FEDER/Junta de Andaluc\'{i}a-Consejer\'{i}a de Transformaci\'{o}n Econ\'{o}mica, Industria, Conocimiento y Universidades.

This work is based on observations made with the NASA/ESA/CSA James Webb Space Telescope. The data were obtained from the Mikulski Archive for Space Telescopes at the Space Telescope Science Institute, which is operated by the Association of Universities for Research in Astronomy, Inc., under NASA contract NAS 5-03127 for JWST. These observations are associated with program \#1199, 1208, 2883, 3362 and 4527.
The specific observations analyzed can be accessed via \dataset[doi:10.17909/18nv-np70]{https://doi.org/10.17909/18nv-np70} (CANUCS/Techinicolor) 
and \dataset[doi:10.17909/h2e3-3t68]{https://doi.org/10.17909/h2e3-3t68} (MAGNIF).
Support for program \#2883 was provided by NASA through a grant from the Space Telescope Science Institute, which is operated by the Association of Universities for Research in Astronomy, Inc., under NASA contract NAS 5-03127.

This research is based on observations made with the NASA/ESA Hubble Space Telescope obtained from the Space Telescope Science Institute, which is operated by the Association of Universities for Research in Astronomy, Inc., under NASA contract NAS 5–26555. These observations are associated with program(s) 13504 and 15117.
The specific observations analyzed can be accessed via \dataset[doi:10.17909/t9-w6tj-wp63]{https://doi.org/10.17909/t9-w6tj-wp63}.

\end{acknowledgments}

\software{Astropy \citep{Astropy2022},
CIGALE \citep{Burgarella2005, Noll2009, Boquien2019}, 
PHOTUTILS \citep{Bradley2025},
STPSF \citep{Perrin14}, 
PySersic \citep{PashaMiller2023},
EAZY \citep{Brammer2008}.}
% pixedfit \citep{AL2021},
% pixedfitbin \citep{AA2017}}.

\appendix
\section{SED fitting parameters}
\label{SED_table}
We present a summary Table \ref{SED_para_table} of the input parameters for \texttt{CIGALE} SED fitting.
% \begin{figure}
%     \centering
%     \includegraphics[width = \columnwidth]{figures/clump_sed.png}
%     \caption{}
%     \label{sed_pz_clumps}
% \end{figure}

\begin{table}[]
    \centering
    \caption{Input parameters for the \texttt{CIGALE} SED fitting.}
    \label{SED_para_table}
    % \resizebox{\columnwidth}{!}{
    \begin{tabular}{cc}
        \hline
        Parameters & Values \\
        \hline
        \multicolumn{2}{c}{sfhdelayed}\\
        \hline
         $\tau$ & 50, 100, 200, 300, 500, 700, 1000, 1500, 2000, 3000, 5000 (Myr) \\
        Age & 20, 30, 50, 80, 120, 160, 200, 250, 300, 400, 500, 600, 700, 800, 900, 1000, 1050(Myr)\\
        $f_{\rm burst}$ & 0.0 \\
        \hline
        \multicolumn{2}{c}{bc03} \\
        \hline
        imf & 1 (Chabrier) \\
        metallicity & 0.02 \\
        seperation\_age & 10 Myr\\
        \hline
        \multicolumn{2}{c}{nebular} \\
        \hline
        logU & -3.5, -3.0, -2.5, -2.0, -1.5, -1.0\\
        zgas & 0.02\\
        ne & 100 \\
        $f_{\rm esc} $ & 0.0 \\
        $f_{\rm dust} $ & 0.0 \\
        lines\_width & 300.0 $\rm km\,s^{-1}$ \\
        \hline
        \multicolumn{2}{c}{dustatt\_calzleit} \\
        \hline
        E\_BVs\_young & 0, 0.05, 0.1, 0.15, 0.2, 0.25, 0.30, 0.45, 0.6, 0.8\\
        E\_BVs\_old\_factor & 0.44\\
        uv\_bump\_amplitude & 0.0\\
        powerlaw\_slope & 0.0\\
        filters & galex.FUV \& generic.bessell.B \& generic.bessell.V \\
        \hline
    \end{tabular}

    % }
\end{table}

% \begin{table}[!ht]
%     \centering
%     \caption{\texttt{piXedfit}}
%     \label{pixedfit_para_table}
%     % \resizebox{\columnwidth}{!}{
%     % \begin{threeparttable}
%     \begin{tabular}{cc}
%         \hline
%         Parameters & Range \\
%         \hline
%         $\log Z/Z_\odot$ & [-2.0, 0.2] \\
%         $\log\tau(\rm Gyr)$ & [-2.0, 1.5]\\
%         $\log\rm age (\rm Gyr)$ & [-2.0, 0.1]\\
%         $\tau_{V}^{\rm ISM}$ & 0.0, 4.0]\\
%         $\log U$ & [-4.0, -2.0]\\
%         \hline
%     \end{tabular}
%     % }
%     % \begin{tablenotes}
%     %   \item [a] 弥散星际介质（Diffuse Interstellar Medium, Diffuse ISM）的尘埃光深。
%     % \end{tablenotes}
%     % \end{threeparttable}
% \end{table}

\section{SED fitting for subcomponents}
\label{app_sed_sub}
\begin{figure*}
    \centering
    \includegraphics[width = 0.95\textwidth]{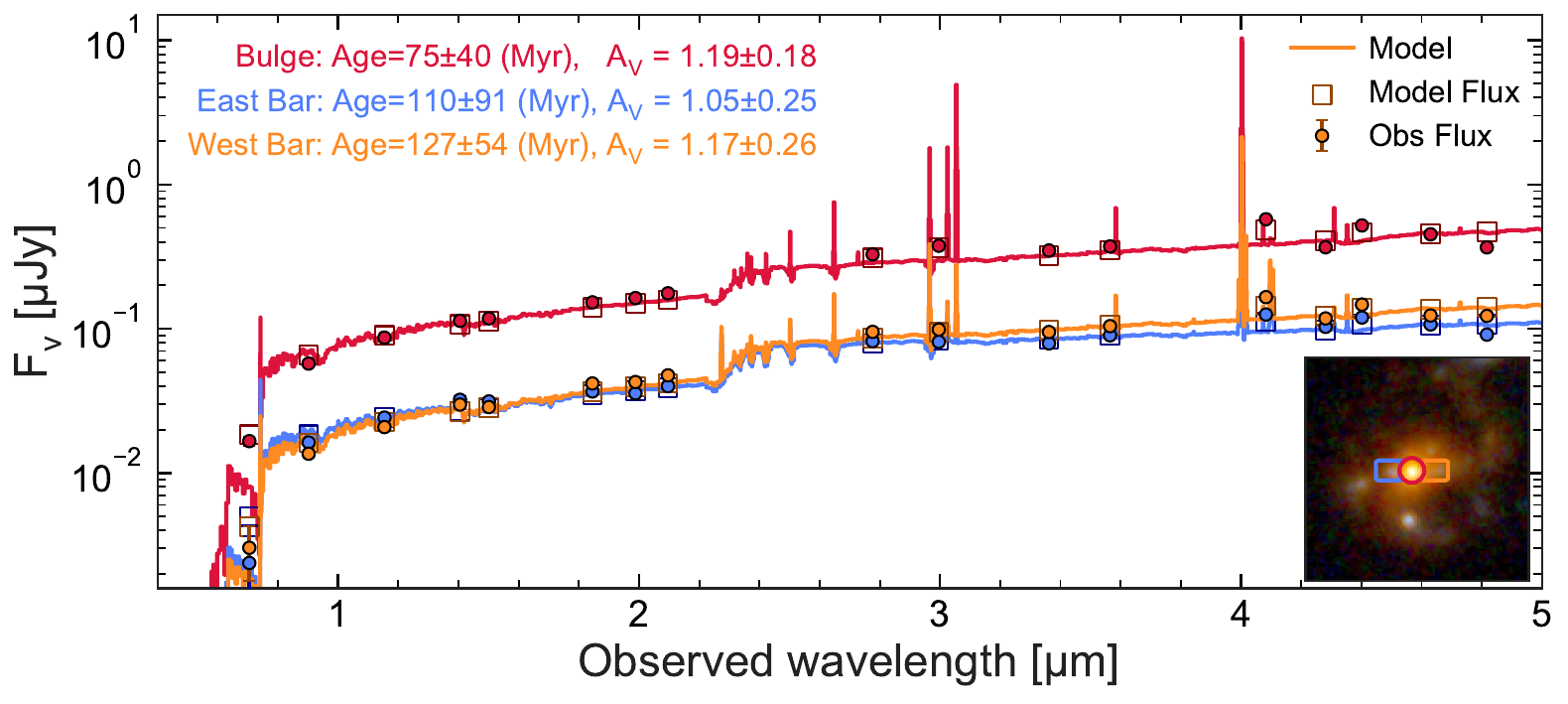}
    \caption{Best-fit CIGALE SED models and JWST photometry for the bulge and the two sides of the bar. The photometry is measured from the apertures shown in the inset, with all images PSF-matched to F480M. The bulge, east bar, and west bar are shown in red, blue, and orange. Observed photometries are shown in filler circles, and best-fit model fluxes for different filters are shown in squares. The east bar is slightly bluer than the west bar, consistent with its younger best-fit age and lower dust attenuation.}
    \label{sed_comps}
\end{figure*}

We apply SED fitting to three subregions, the central bulge and the east and west sides of the bar, using the apertures shown in Fig.~\ref{sed_comps}. The fitting is performed with CIGALE under the same assumptions as the integrated SED fitting (Table~\ref{SED_para_table}).
The SED fitting derives stellar age of $75\pm40$ Myr, SFR of $39\pm12\, \rm M_\odot/yr$, stellar mass of $10^{9.6\pm0.1}\, \rm M_\odot$ and $A_V = 1.19\pm0.18$ mag for the bulge.
For the east bar, the SED fitting derives stellar age of $110\pm91$ Myr, SFR of $8\pm2\, \rm M_\odot/yr$, stellar mass of $10^{9.0\pm0.1}\, \rm M_\odot$ and $A_V = 1.05\pm0.25$ mag.
For the west bar, the SED fitting derives stellar age of $127\pm54$ Myr, SFR of $10\pm3\, \rm M_\odot/yr$, stellar mass of $10^{9.2\pm0.1}\, \rm M_\odot$ and $A_V = 1.17\pm0.26$ mag.
The east bar is slightly bluer and less obscured, while the west bar is slightly dustier, more massive, and has a slightly higher SFR. Although the differences are modest, the SED-inferred ages of the bar-side regions are consistently around ($\sim100$) Myr, suggesting that the stellar populations associated with the bar-like structure may have already been in place by ($z\sim5.5$).

\section{Emission line properties}
\label{emlines}
\begin{figure*}
    \centering
    \includegraphics[width = \textwidth]{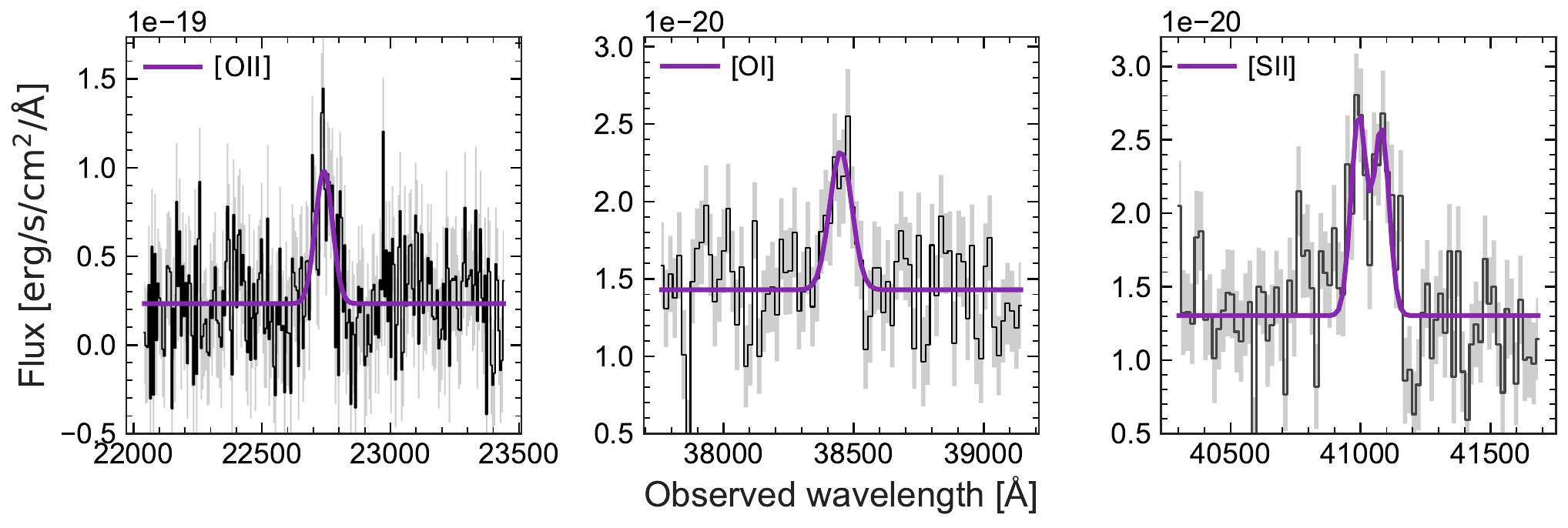}
    \caption{Complementary to the bottom panel of Fig.~\ref{img}. From left to right: spectral regions of emission lines, [O\,{\sc i}], [O\,{\sc ii}] and [S\,{\sc ii}]. The purple curves show the best-fit line models.}
    \label{spec_app}
\end{figure*}

We present a summary Table \ref{line_fluxes} of the emission line fluxes, and spectral regions of emission lines in Fig.~\ref{spec_app} complementary to Fig.~\ref{img}.

\begin{deluxetable}{cc}
    \tablecaption{Rest-optical emission line fluxes\label{line_fluxes} \tablenotemark{$\dagger$}}
    \tablewidth{\textwidth}
    \tablehead{
    \colhead{Ion+Wavelength (\AA)} & \colhead{Flux ($10^{-19}$ erg s$^{-1}$ cm$^{-2}$)}
    }
    \startdata
    [O {\sc ii}] $\lambda\lambda$3726,3729\tablenotemark{$\ddagger$} & $62.7\pm11.0$\\
    H$\beta$ & $15.5\pm2.0$\\
    \ [O {\sc iii}] $\lambda$5007 & $40.7\pm3.4$\\
    \ [O {\sc i}] $\lambda$6302 & $9.5\pm1.9$\\
    H$\alpha$ (narrow) & $57.7\pm4.6$\\
    H$\alpha$ (broad) & $106.3\pm9.4$\\
    \ [N {\sc ii}] $\lambda$6584 & $39.1\pm3.6$\\
    \ [S {\sc ii}] $\lambda$6716 & $10.0\pm1.4$\\
    \ [S {\sc ii}] $\lambda$6731 & $9.3\pm1.4$\\
    \enddata
    \tablenotetext{}{$\dagger$ The spectra were extracted using the default \texttt{msaexp} procedure including its path-loss correction.We further compare the spectra with the photometry over the corresponding wavelength ranges, obtaining scale factors of $\sim 1.7$ for G140M and $\sim 2.7$ for G395M. 
    The fluxes listed in this table are the original \texttt{msaexp} measurements and have not been corrected by these additional rescaling factors. The narrow-line fluxes may therefore be underestimated by the corresponding factors.}
    \tablenotetext{}{$\ddagger$ The doublet is not resolved, and a single Gaussian is fit to the blended line.}

\end{deluxetable}

\bibliography{sample701}{}
\bibliographystyle{aasjournalv7}

\end{document}